%% file: hydro-IIb.tex
\documentclass{aa}
\usepackage[varg]{txfonts}
\usepackage{graphicx}
\usepackage{natbib}
\usepackage{booktabs}
\usepackage{morefloats}
\bibpunct{(}{)}{;}{a}{}{,}

\begin{document}


\title{Hydrodynamical modelling of Type IIb SNe.\thanks{This manuscript is based on preliminary data from the Carnegie Supernova Project (CSP), and was included in the PhD thesis of M. Ergon (2015). The thesis introduction is available in electronic form from the Swedish digital science archive (DIVA; http://www.diva-portal.org/smash/get/diva2:800078/FULLTEXT01.pdf).}}

\author{M.~Ergon\inst{\ref{inst1}} \and M.~Stritzinger\inst{\ref{inst2}} \and F.~Taddia\inst{\ref{inst1}} \and J.~Sollerman\inst{\ref{inst1}} \and C.~Fransson\inst{\ref{inst1}}}

\institute{The Oskar Klein Centre, Department of Astronomy, AlbaNova, Stockholm University, 106 91 Stockholm, Sweden 
\label{inst1}
\and Department of Physics and Astronomy, Aarhus University, Ny Munkegade 120, 8000 Aarhus C, Denmark
\label{inst2}}

\date{To be submitted to Astronomy and Astrophysics.}

\abstract{We present HYDE, a new one-dimensional hydrodynamical code, and use it to construct a grid of supernova (SN) models based on solar-metallicity bare helium-core models evolved to the verge of core-collapse with MESA STAR. This grid is suited to model Type IIb SNe, which progenitor stars are thought to have lost all but a tiny fraction of their hydrogen envelopes. As previously demonstrated, such an envelope only affects the early lightcurve, and the diffusion phase and the early tail phase lightcurves are governed by the helium core. Relatively massive hydrogen envelopes do, however, affect the photospheric velocities during the diffusion phase, which could lead to underestimates of the explosion energy. Using an automated procedure we fit the bolometric lightcurves and photospheric velocities for a large sample of (17) Type IIb SNe to the grid of SN models. We find that the distribution of initial masses for the sample can be reasonably well described by a standard Salpeter IMF, although there is an under-population in the >25 M$_\odot$ range. The fractions of SNe with initial masses <15 M$_\odot$ and <20 M$_\odot$ are 56 and 81 percent, respectively, suggesting either the binary channel to dominate the production of Type IIb SNe or a 
flaw in our understanding of single-star mass-loss. We find correlations between the explosion energy, initial mass and mass of \element[ ][56]{Ni}; the explosion energy increases with initial mass and the mass of \element[ ][56]{Ni} increases with explosion energy. The method used allows us to determine the errors in the model parameters arising from the observed quantities and the degeneracy of the solution. We find that an error in the distance and extinction propagates mainly to the derived mass of \element[ ][56]{Ni}, whereas an error in the photospheric velocity propagates mainly to the derived helium-core mass and explosion energy. Fits using the bolometric lightcurve alone are completely degenerate along the M$_{\mathrm{ej}}^{2}$/E$_{\mathrm{ej}}$=const curve, whereas fits using also the photospheric velocities are quite robust for well-sampled SNe. Finally, we provide a description and tests of the HYDE code, and a discussion of the limitations of the method used.}


\titlerunning{Hydrodynamical modelling of Type IIb SNe.}
\authorrunning{M. Ergon et al.}
\maketitle

\defcitealias{Erg14a}{E14a}
\defcitealias{Erg14b}{E14b}
\defcitealias{Ber12}{B12}
\defcitealias{Jer14}{J14}
\defcitealias{Str15}{S15}

\section{Introduction}

Type IIb supernovae (SNe) are thought to arise from stars that have lost most of their hydrogen envelopes, either through stellar winds or through Roche-lobe overflow to a binary companion. These SNe are observationally characterized by a transition from Type II (with a hydrogen signature) to Type Ib (without a hydrogen but with a helium signature). Whether binary or single progenitor systems are dominating the production of Type IIb SNe is still debated, but for SN 1993J a companion star has been detected by direct observations \citep{Mau04,Fox14}. Because most of the hydrogen envelope has been lost, whereas the helium core is still intact, we expect these SNe to be well approximated by the explosions of bare helium cores, except during the early cooling phase. This method has been used by \citet{Ber12}, and allows for estimates of the helium-core mass, explosion energy and mass of \element[ ][56]{Ni}, whereas the progenitor radius can not be estimated without taking the hydrogen envelope into account. Type IIb SNe have the unique quality to allow an estimate of the helium-core mass which, contrary to the ejecta mass, is directly linked to the initial mass of the star. Some parameter studies have been published \citep[e.g.][]{Lym14}, but are all based on approximate lightcurve modelling \citep[e.g.][]{Arn82}. The aim of this paper is to use the new hydrodynamical code HYDE to construct a grid of SN models based on bare helium-core models, and use this to estimate the progenitor and SN parameters for a large sample of Type IIb SNe. The sample consists of the Type IIb SNe from the Carnegie Supernova Project (CSP) sample of stripped-envelope SNe (Stritzinger~et~al.~2015), 
as well as most of the Type IIb SNe that have been individually studied in the literature.

Application of hydrodynamics to SNe lightcurves was introduced in the 70ths \citep[e.g.][]{Fal77}, and since then a number of codes spanning a wide range of complexity have followed. Some implements more advanced physics, as multi-dimensional \citep[e.g.][]{Mul91} and radiation \citep[e.g.][]{Bli98} hydrodynamics, whereas others are one-dimensional and based on the diffusion approximation \citep[e.g.][]{Ber11}. The different codes all have their different applications and no code is yet capable of modelling a core-collapse (CC) SN consistently, including all the relevant physics. HYDE belongs to the latter category, and like other simplified codes it has the benefit of being fast, which is critical when building model grids covering large volumes of parameter space. The use of model grids to determine the progenitor and SN parameters has been explored before \citep[e.g.][]{Lit83,Lit85} with somewhat mixed results \citep[e.g.][]{Ham03}, but the decreasing computational cost and the increasing amount of data, motivate a renewed interest in this approach. A model grid also allows the degeneracy of the solution and the errors in the progenitor and SN parameters to be estimated.

The paper is organized as follows. In Sect.~\ref{s_hyde} we describe and test the HYDE code. In Sect.~\ref{s_model_grid} we describe the grid of bare helium-core and SN models, and discuss the dependence of the observed properties on the progenitor and SN parameters. In Sect.~\ref{s_hydrogen_envelope} we present models with low-mass hydrogen envelopes, and discuss the effects of these on the observed properties. In Sect.~\ref{s_modelling_IIb} we describe our fitting procedure, use the grid of SN models to estimate the progenitor and SN parameters for our sample of Type IIb SNe, and discuss the total sample statistics. The observational details for the Type IIb SNe from the CSP sample and those individually studied in the literature are given in Appendices~\ref{a_csp_sample} and \ref{a_literature_sample}, respectively. Finally, we conclude and summarize the paper in Sect.~\ref{s_conclusions}. 

\section{The HYDE code}
\label{s_hyde}

HYDE is a 1-D (spherically symmetric) hydrodynamical code based on the diffusion approximation, developed along the lines described in \citet{Fal77}. The code is written in C++ and may also be run in homologous mode, where the dynamics has been switched off and the thermodynamical state is solved for given the constraint of homologous expansion. The code is also configurable in a number of other ways, e.g.~with respect to the use of a flux-limiter and the form of the momentum and energy equations, and atomic and opacity data are read from files in generic (but proprietary) formats, and can therefore easily be updated.

\subsection{Hydrodynamics}
\label{s_hyde_hydro}

HYDE solves the hydrodynamical conservation equations for mass, momentum and thermal energy, coupled with the diffusion approximation for the radiation field 
(eqs.~1-4, \citealp{Fal77}). 
To limit the diffusion velocity in the optically thin regime, HYDE provides an option to use a flux limiter following the prescription by \citet{Ber11}, which is then added in the diffusion equation (eq.~4, \citealp{Fal77}). The flux limiter transforms the radiation field from the optically thick diffusion limit to the free-streaming unidirectional limit, but is only qualitatively correct in the intermediate region \citep{Mih84}, and may introduce inconsistencies in the radiation field \citep{Eps81}. These inconsistencies arise in the radiation pressure terms in the momentum and thermal energy equations, as the flux-limiter enforces a transformation from an isotropic radiation field to an unidirectional one, i.e.~the Eddington factor ($f_\mathrm{E}=P_\mathrm{R}/E_\mathrm{R}$) increases from 1/3 to 1. Therefore, HYDE provides an option to use alternative forms of the momentum and thermal energy equations, modified to be consistent with the flux limiter. To make the momentum equation consistent, we add the isotropy related factor $(3P_\mathrm{R}-E_\mathrm{R})/r$ \citep[eq~.96.3]{Mih84}, and use the time-independent first-order moment equation to 
express 
the divergence of the radiation pressure in terms of radiative energy flux. To make the thermal energy equation consistent, we add the isotropy related factor $(3P_\mathrm{R}-E_\mathrm{R})v/(\rho r)$ \citep[e.q.~96.9]{Mih84}, and use the time-independent first-order moment equation to determine the Eddington factor for a radiation field with the energy density of a blackbody and the flux given by the flux limiter. This, in turn, is used to 
express 
the radiation pressure in terms of radiation energy density and the Eddington factor, and we arrive at the following modified versions of eqs.~1-4 in \citet{Fal77}

\begin{equation}
\label{eqrhyd1}
m = \int 4 \pi r^{2} \rho dr
\end{equation}

\begin{equation}
\label{eqrhyd2}
{{\partial v} \over {\partial t}} = - 4 \pi r^{2} {{\partial P_\mathrm{G}} \over {\partial m}} + {{\rho \kappa_\mathrm{F}} \over {4 \pi r^{2} c}} L + {{G m} \over {r^{2}}}
\end{equation}

\begin{equation}
\label{eqrhyd3}
{{\partial} \over {\partial t}} \left(E_{\mathrm{G}}+{{E_\mathrm{R}}\over {\rho}} \right) = - (P_\mathrm{G}+f_\mathrm{E} E_\mathrm{R}) {{\partial} \over {\partial t}} \left({{1}\over {\rho}}\right)+ (3f_\mathrm{E}-1) E_\mathrm{R} {{v} \over {\rho r}} - {{\partial L} \over {\partial m}}  + \epsilon
\end{equation}

\begin{equation}
\label{eqrhyd4}
L=-(4 \pi r^{2})^{2} {{\lambda a c} \over {3 \kappa_\mathrm{R}}}  {{\partial T^{4}}  \over {\partial m}}
\end{equation}

\begin{equation}
\label{eqrhyd5}
{{\partial (f_\mathrm{E} E_\mathrm{R})} \over {\partial r}} + (3f_\mathrm{E}-1){{E_\mathrm{R}} \over {r}} = - {{\rho \kappa_\mathrm{E}} \over {4 \pi r^{2} c}} L
\end{equation}

where $E_\mathrm{G}$ and $P_\mathrm{G}$ are the specific (thermal) energy and the pressure for the gas, $E_\mathrm{R}$ and $L$ the energy density and the luminosity for the radiation field, $\lambda$ the flux limiter, $\epsilon$ the specific radioactive heating rate, and $\kappa_\mathrm{R}$ and $\kappa_\mathrm{F}$ the Rosseland mean opacity and the mean opacity weighted in energy flux. 
Note that the Eddington factor is determined by the form of the flux limiter, and may differ from what would be obtained from geometrical considerations

The equation of state (EOS) for the radiation field is assumed to be that of blackbody radiation, although the Eddington factor is allowed to vary between 1/3 and 1 as described. The EOS for the gas is assumed to be that of an ideal gas, including the effects of ionization, but excluding those of excitation, and degeneracy of the electron gas is not taken into account. The ion
and 
electron number densities required in the EOS are calculated by solving the Saha equation, using up to five ionization stages for each element (as configured). 
To solve the equations,  
initial and boundary conditions
also need to be specified. 
At the inner boundary we adopt $L=0$ and $v=0$, and at the outer boundary $P_\mathrm{G}=0$ and $T^{4}=(3/4)T_\mathrm{\text{eff}}^4(\tau+q)$. The value of $q$ may be set to 2/3 or 1/3, which gives the Eddington approximation or a modified version of it, chosen to be consistent with the flux limiter in the sense that the unidirectional limit ($f_\mathrm{E}=1$) is recovered when $\tau \rightarrow 0$. The initial conditions are determined by the stellar model, and for consistency the temperature structure of the model may be recalculated using the HYDE EOS assuming hydrostatic equilibrium. The zero-velocity boundary represents the division (mass cut) between the ejected material and the compact remnant, and the explosion energy is injected near this boundary in the form of thermal energy (thermal bomb), represented as an additional heating term in Eq.~\ref{eqrhyd3}.

Given the initial and boundary conditions, the EOS, the opacity (Sect.~\ref{s_hyde_opacity}) and the radioactive heating (Sect.~\ref{s_hyde_radio}), the equation system (Eqs.~\ref{eqrhyd1}-\ref{eqrhyd4}) is solved by a finite difference scheme similar to the one described by \citet[eqs.~A1-A12]{Fal77}. To handle strong velocity gradients (shocks) an artificial viscosity following the prescription by \citet{Neu50} is used, and added to the pressure terms in the momentum and thermal energy equations (Eqs.~\ref{eqrhyd2} and \ref{eqrhyd3}). The dynamical state is solved for using the momentum equation (Eq.~\ref{eqrhyd2}) and a forward difference scheme, where the new state is explicitly determined by the previous state, and is therefore trivial to advance. The thermodynamical state is solved for using the thermal energy equation (Eq.~\ref{eqrhyd3}) and a backward difference scheme, where the new state is implicitly determined by the previous state. This results in a non-linear equation system, which is solved by a Newton-Raphson like method, where the equation system is linearised in terms of temperature corrections \citep[][appendix AIa]{Fal77}. The time-step $\Delta t$ for each calculation is initially set according to the Courant-Friedrich-Levy (CFL) condition $v_\mathrm{s}\Delta t/ \Delta r=0.5$, where $v_\mathrm{s}$ and $\Delta r$ are the sound-speed and the radial size of the cell, respectively, and is subsequently reduced if the (configurable) convergence criteria are not satisfied. 

\subsection{Opacity}
\label{s_hyde_opacity}

The Rosseland mean opacity, the sole quantity that determines the coupling between the matter and the radiation field in the diffusion approximation, is interpolated from the OPAL opacity tables \citep{Igl96} complemented with the low temperature opacities by \citet{Ale94}. Note, however, that in the modified energy equation (Eq.~\ref{eqrhyd5}) also the mean opacity weighted in radiative energy flux appears, but we will assume the difference to be small, as is justified if the (grey) electron scattering opacity dominates. The opacity tables are calculated for a static medium in LTE, and therefore the line opacity as well as the opacity in the optically thin region may be underestimated (Sect.~\ref{s_model_limitations}). To compensate for this lack of opacity, we use a minimum value for the opacity, commonly referred to as opacity floor. The value of this opacity floor is set to $0.01$ cm$^2$ g$^{-1}$ in the hydrogen envelope and $0.025$ cm$^2$ g$^{-1}$ in the helium core, following \citet[private communication]{Ber12}, who calibrated these values by comparison to the STELLA hydrodynamical code \citep{Bli98}. 

\subsection{Radioactive heating}
\label{s_hyde_radio}

The transfer of the $\gamma$-rays and positrons emitted in the decay chains of \element[ ][56]{Ni}, \element[ ][57]{Ni} and \element[ ][44]Ti is calculated with a Monte-Carlo method similar to that by \citet{Jer11,Jer12}, and the mass fractions of the isotopes evolved at each time step. The grey opacities, luminosities and decay times used are the same as in \citet{Jer11,Jer12}. 
The deposited decay luminosity is assumed to contribute only to the heating of the gas, which is a fair approximation in the optically thick region where the degree of ionization is high \citep{Koz92}, and the heating rate is fed into the energy equation.

\subsection{Observed luminosity}
\label{s_observed_luminosity}

The observed luminosity may be taken as the comoving frame luminosity at the outer boundary, but as this boundary could be accelerated to high speeds, a transformation to the observer frame may be necessary, and the light-travel time may need to be taken into account. Therefore, HYDE provides an option to calculate the observed luminosity as $L_\mathrm{\text{obs}}(t_\mathrm{\text{obs}})=[1+2\beta(t)]L(t)$, where $\beta(t)=v(t)/c$ and $t_\mathrm{\text{obs}}=t-v(t)/c$. This expression assumes a free-streaming unidirectional radiation field, and can be derived from first principles (see e.q~99.39 in \citealp{Mih84}). HYDE also provides an option to proceed from the luminosity at some inner surface, outside which the optical depth and the radioactive energy deposition is negligible. This reduces the influence of the flux limiter, and in this case the difference between the luminosity in the comoving frame and that measured by a distant observer is less critical.

\subsection{Tests of the code} 

\begin{figure}[tbp!]
\includegraphics[width=0.5\textwidth,angle=0]{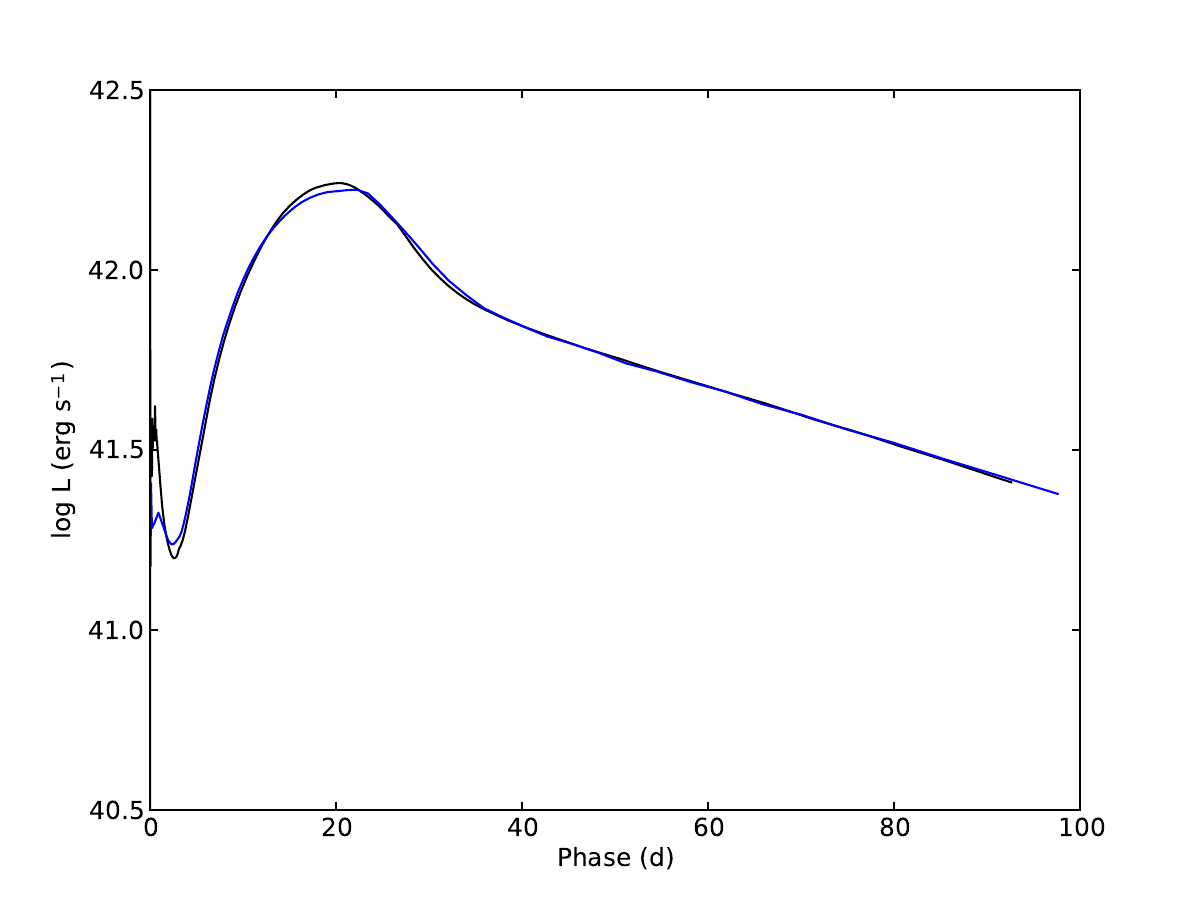}
\caption{Bolometric lightcurve for the 4 M$_\odot$ bare helium-core model from \citet{Nom88} as modelled with HYDE (black) and the adjusted version of the \citet{Ber12} He4 model presented in \citetalias{Erg14a} (blue).}
\label{f_b12_model_comp}
\end{figure}

\begin{figure}[tbp!]
\includegraphics[width=0.5\textwidth,angle=0]{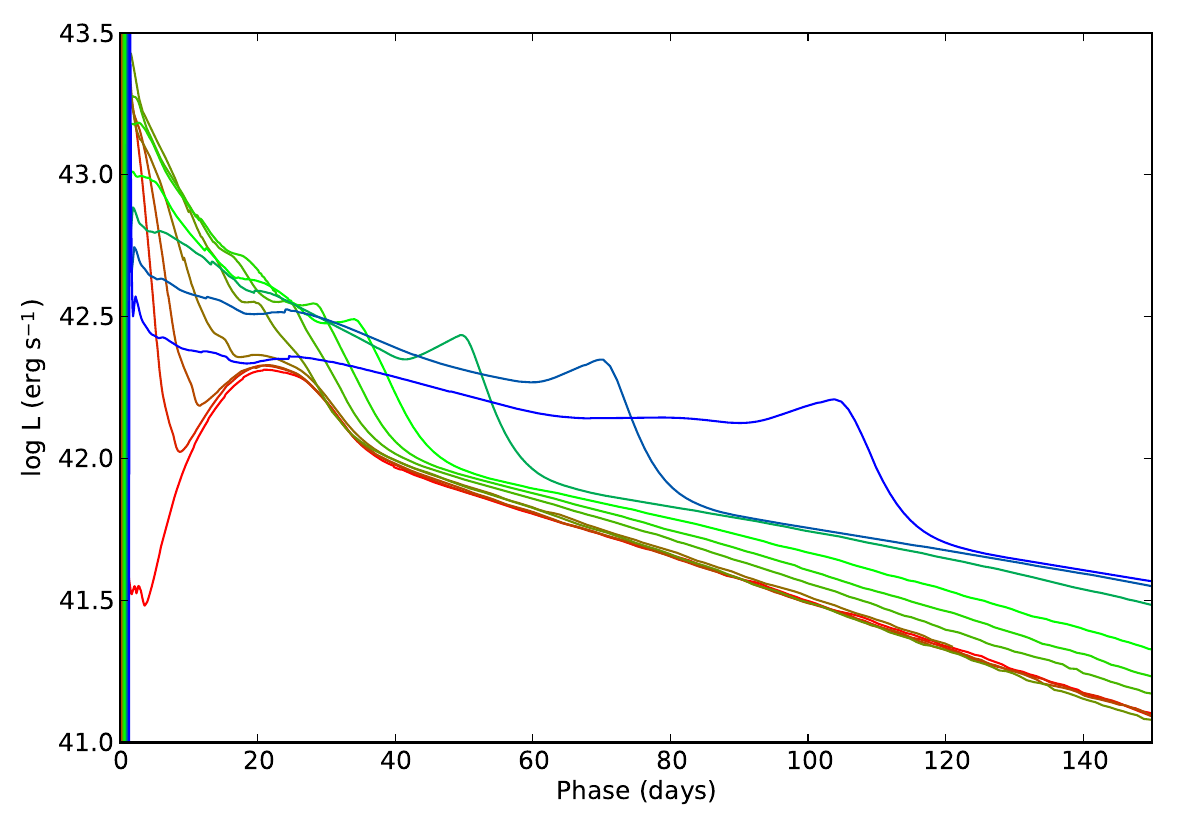}
\caption{Progression of model bolometric lightcurves calculated with HYDE for a 15 M$_\odot$ MESA model with the mass-loss adjusted to yield a final mass of 11.0, 8.0, 6.0, 5.0, 4.8, 4.6, 4.4, 4,2, 4.1, 4.05 and 4.0 M$_\odot$, colour coded from blue (11.0 M$_\odot$) to red (4.0 M$_\odot$). The explosion parameters were E=1.0$\times$10$^{51}$erg, M$_{\mathrm{Ni}}$=0.1 M$_\odot$ and Mix$_{\mathrm{Ni}}$=M$_{\mathrm{He}}$/M (Sect.~\ref{s_model_grid}).}
\label{f_M15_massloss}
\end{figure}

\begin{figure}[tbp!]
\includegraphics[width=0.5\textwidth,angle=0]{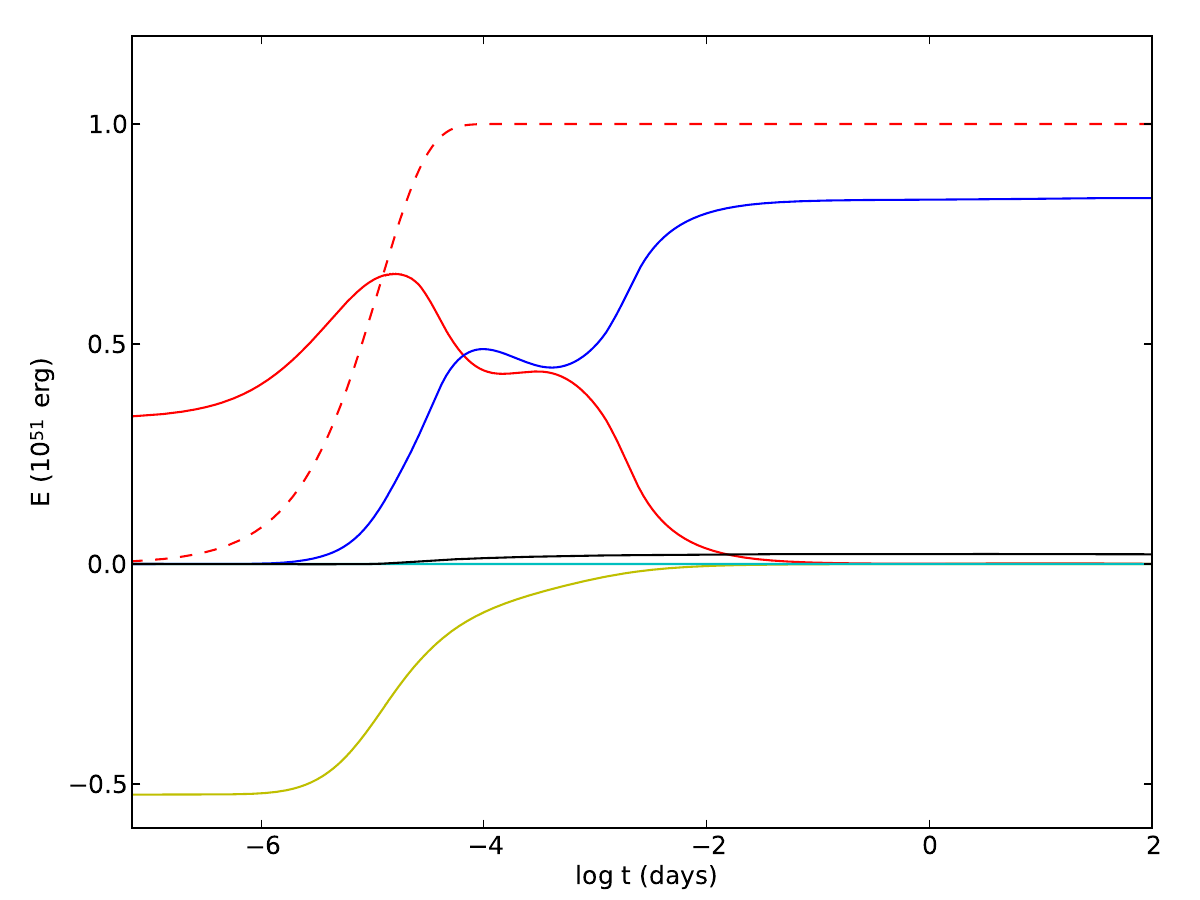}
\caption{The change in the total energy minus the net energy gained (black solid line) calculated with HYDE for a model with M$_\mathrm{He}$=4.0 M$_\odot$, E=1.0$\times$10$^{51}$erg, M$_{\mathrm{Ni}}$=0.1 M$_\odot$ and Mix$_{\mathrm{Ni}}$=1.0, where the net energy gained is given by the sum of the explosion energy, radioactive heating and radiative losses. For comparison we also show the thermal (red solid line), kinetic (blue solid line), gravitational (yellow solid line), ionization (cyan solid line) and injected explosion energy (red dashed line) for this model.}
\label{f_hyde_energy_conserve}
\end{figure}

The homologous behaviour has been tested by comparison to analytical solutions by \citet{Ims92}, and the deposition of radioactive decay energy by comparison to the steady-state NLTE code described in \citet{Jer11,Jer12}. Energy conservation has been tested, and is accurate to a few percent of the explosion energy in a typical run. This is illustrated by Fig.~\ref{f_hyde_energy_conserve}, which shows the change in the total energy minus the net energy gained (sum of explosion energy, radioactive heating and radiative losses), for a model with M$_\mathrm{He}$=4.0 M$_\odot$, E=1.0$\times$10$^{51}$erg, M$_{\mathrm{Ni}}$=0.1 M$_\odot$ and Mix$_{\mathrm{Ni}}$=1.0. This quantity, which should be zero if energy is conserved, is <0.023$\times$10$^{51}$ erg, and for comparison we also show the thermal, kinetic, gravitational, ionization and injected explosion energy for this model. The hydrodynamical behaviour has been tested by comparison to the results presented in \citet{Ber12} using the same 4 M$_\odot$ bare helium-core model from \citet{Nom88}. Figure~\ref{f_b12_model_comp} shows a comparison between the lightcurve calculated with HYDE and the lightcurve for the adjusted version of the \citet{Ber12} He4 model presented in \citet[][hereafter \citetalias{Erg14a}]{Erg14a}. Both models have the same explosion parameters (E=1.0$\times$10$^{51}$ erg, M$_{\mathrm{NI}}$=0.075 M$_\odot$ and Mix$_{\mathrm{Ni}}$=0.95), and for consistency flux-limited diffusion without a calculation of the Eddington factor was used, and only ionization of hydrogen and helium was included in the EOS. Except at $\lesssim$1 day the 
bolometric lightcurves are very similar, 
but the luminosity during the first day is considerably higher in our model.
The reason for this could be differences in the zoning or the density profile, and as we have scanned the model from \citet{Nom88} such differences are expected. Figure~\ref{f_M15_massloss} shows lightcurves calculated with HYDE for a series of 15 M$_\odot$ MESA models where the mass-loss was adjusted to yield final masses in the range 11$-$4 M$_\odot$. The sequence of lightcurves shows the expected transformation from an explosion energy powered Type IIP like lightcurve to a radioactively powered Type Ib like lightcurve. Further justification for a healthy behaviour is provided in Sects.~\ref{s_model_grid} and \ref{s_hydrogen_envelope}, where we discuss the observed and physical properties of our bare helium-core and extended models, respectively.

\section{The model grid}
\label{s_model_grid}

\subsection{Progenitor models}
\label{s_prog_models}

The progenitor models were constructed using MESA STAR \citep{Pax11,Pax13} by evolving solar-metallicity helium cores until the verge of core-collapse. This is similar to what was done in \citet{Nom88}, and relies on the assumption that the hydrogen envelope does not appreciably affect the evolution of the helium core. Evolving a set of solar-metallicity 15 M$_\odot$ models, 
where we adjusted 
the mass loss to yield final masses in the range 15$-$4 M$_\odot$, resulted in helium cores of similar size and composition, in support of this assumption. All models are non-rotating and the Schwarzschild criteria is used for convection. Otherwise, the default MESA configuration is used, and the evolution was terminated at a central density of 10$^{9.5}$ g cm$^{-3}$, which typically occurred slightly before core-collapse. The evolved models spans M$_{\mathrm{He}}$=4.0-5.0 M$_\odot$ in 0.25 M$_\odot$ steps and M$_{\mathrm{He}}$=5.0-10.0 M$_\odot$ in 0.5 M$_\odot$ steps, and in Table \ref{t_mg_mesa_model} we give the helium, carbon-oxygen and iron core masses, the radii and the total (gravitational plus thermal) energy for these models. Below 4.0 M$_\odot$ the late burning stages ignited off centre, which caused convergence problems, and these stellar models were constructed by scaling of the 4.0 M$_\odot$ density profile.

\begin{table}[tb!]
\caption{The helium, carbon-oxygen and iron core masses, the radii and the total (gravitational plus thermal) energy of the progenitor models.}
\begin{center}
\include{mg-mesa-model-table}
\end{center}
\label{t_mg_mesa_model}
\end{table}

\subsection{SN models}
\label{s_sn_models}

HYDE does not include a treatment of 
the core-collapse itself. Instead the outcome of this event is simulated by the injection of thermal energy (thermal bomb) at some location assumed to correspond to the division between the collapsing core and the ejected material. This location is fixed to 1.5 M$_\odot$ in all our models, and the explosion energy (E) is treated as a free parameter. HYDE does not include a network of nuclear reactions, so the explosive nuclear burning in the iron core and the inner parts of the oxygen zones, synthesizing the radioactive isotopes powering the lightcurve, can not be modelled. Because of this, and the absence of multi-dimensional effects as macroscopic mixing in 1-D (spherically symmetric) modelling, the mass (M$_{\mathrm{Ni}}$) and mixing (Mix$_{\mathrm{Ni}}$) of the \element[ ][56]{Ni} are also treated as free parameters. The mass fraction of \element[ ][56]{Ni} ($\mathrm{X}_{\mathrm{Ni}}$) 
is 
assumed to be a linearly declining function of the ejecta mass ($\mathrm{m}_{\mathrm{ej}}$) becoming zero at some fraction (Mix$_{\mathrm{Ni}}$) of the total ejecta mass, expressed as $\mathrm{X}_{\mathrm{Ni}} \propto 1-\mathrm{m}_{\mathrm{ej}}/(\mathrm{Mix}_{\mathrm{Ni}} \mathrm{M}_{\mathrm{ej}}), \mathrm{X}_{\mathrm{Ni}} \geq 0$. Note that this expression allows Mix$_{\mathrm{Ni}}>1$, although the interpretation of the parameter then becomes less clear. The SN explosion is thus parametrized using three parameters (E, M$_{\mathrm{Ni}}$ and Mix$_{\mathrm{Ni}}$) and the progenitor star using one (M$_{\mathrm{He}}$). The total parameter space spanned is M$_{\mathrm{He}}$=2.5-10 M$_\odot$, E=0.4-6.0$\times$10$^{51}$ erg, M$_{\mathrm{Ni}}$=0.015-0.3 M$_\odot$ and Mix$_{\mathrm{Ni}}$=0.5-1.4 using a 21$\times$24$\times$15$\times$9 grid\footnote{The M$_\mathrm{He}$>7 M$_\odot$ and E>2.2$\times$10$^{51}$ erg models are not yet finalized and for SN 1996cb, 2003bg, 2011ei and 2011fu we use an older version of the grid based on (mass) scaled versions of the 4.0 M$_\odot$ bare helium-core model from \citet{Nom88}}. We find this resolution to be sufficient to safely interpolate intermediate values. HYDE was configured to run with the flux-limiter and the modified momentum equation, but without calculation of the Eddington factors. The modified Eddington approximation was used at the outer boundary, and the ionization energy was not included in the EOS. The luminosity was taken from an inner surface, outside which the optical depth and the fractional radioactive energy deposition were <1 percent, and the transformation to the observer frame was ignored\footnote{Some of these settings will be changed for the final version of the model grid, but this should not have any significant effect on the results.}.

\subsection{Dependence on progenitor and SN parameters}
\label{s_dependence}

\begin{figure}[tbp!]
\includegraphics[width=0.5\textwidth,angle=0]{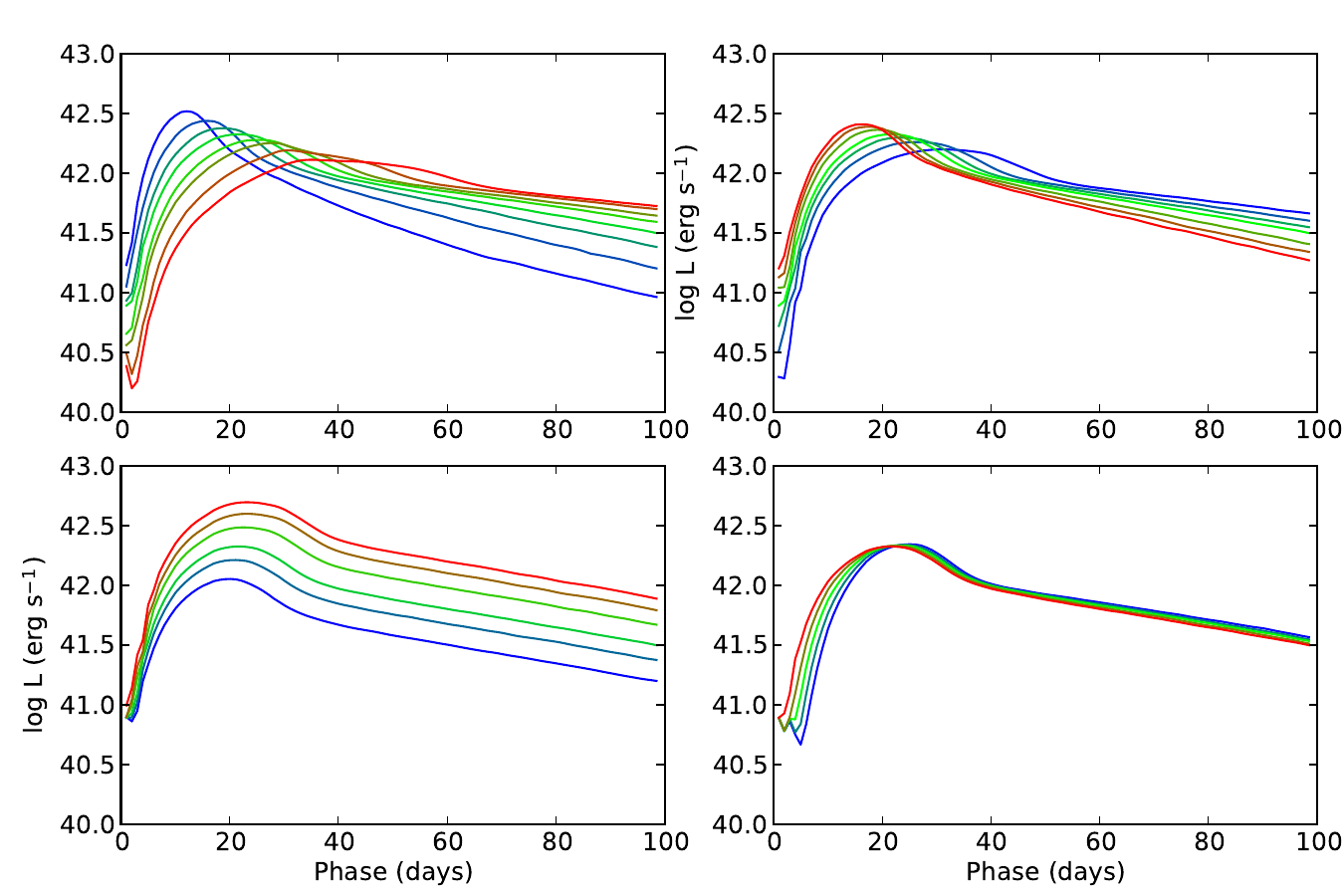}
\caption{Model bolometric lightcurves for day 1-100 showing the dependence on M$_{\mathrm{He}}$ (2.5-7.0 M$_\odot$; upper left panel), E (0.4-2.2$\times$10$^{51}$ erg; upper right panel), M$_{\mathrm{Ni}}$ (0.05-0.25 M$_\odot$; lower left panel) and Mix$_{\mathrm{Ni}}$ (0.6-1.0; lower right panel). Low to high values are displayed in blue to red colour coding and the values for the parameters not varied are M$_{\mathrm{He}}$=4.0 M$_\odot$, E=1.0$\times$10$^{51}$ erg, M$_{\mathrm{Ni}}$=0.1 M$_\odot$ and Mix$_{\mathrm{Ni}}$=1.0.}
\label{f_mg_L}
\end{figure}

\begin{figure}[tbp!]
\includegraphics[width=0.5\textwidth,angle=0]{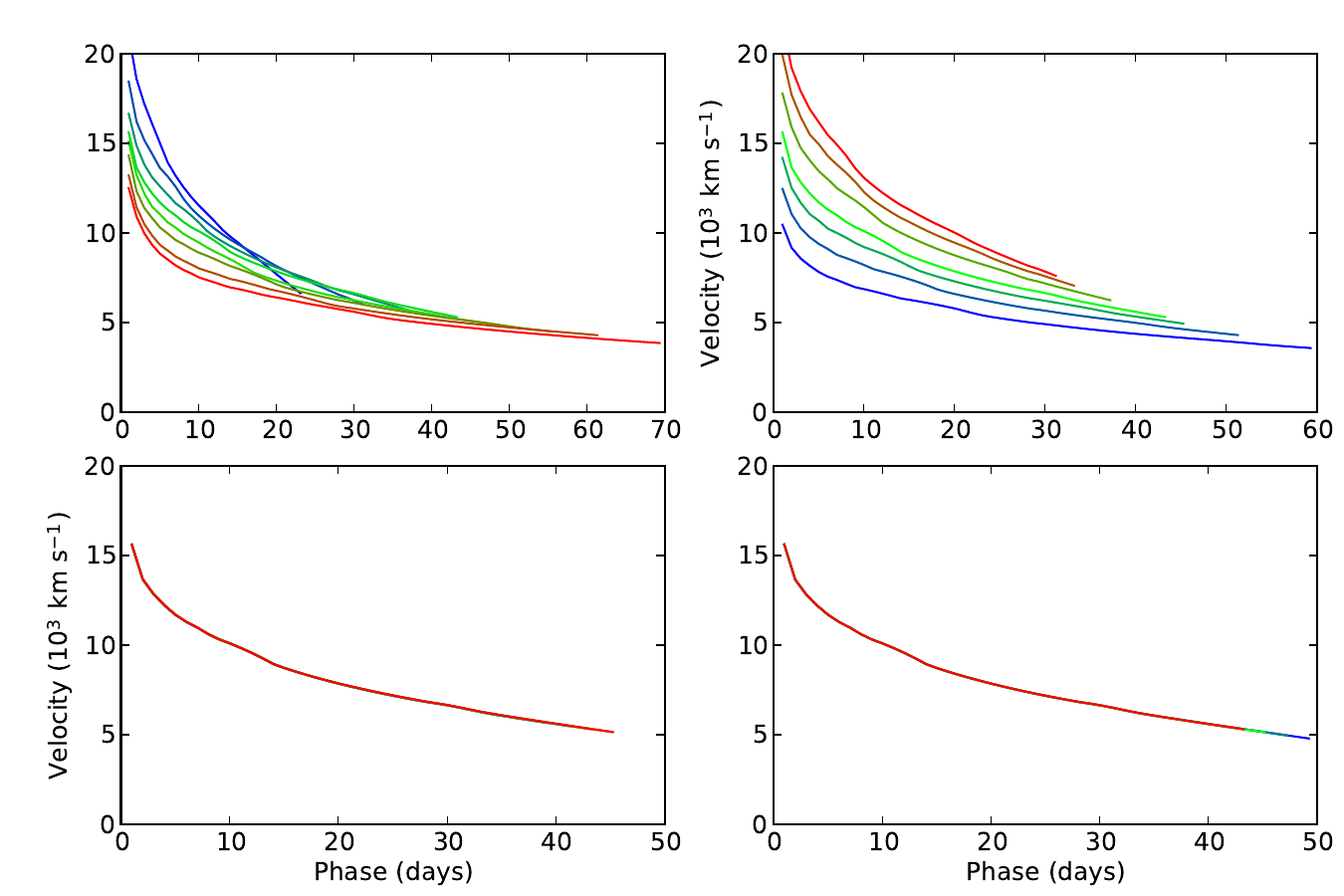}
\caption{Model photospheric velocities for day 1-100 showing the dependence on M$_{\mathrm{He}}$ (2.5-7.0 M$_\odot$; upper left panel), E (0.4-2.2$\times$10$^{51}$ erg; upper right panel), M$_{\mathrm{Ni}}$ (0.05-0.25 M$_\odot$; lower left panel) and Mix$_{\mathrm{Ni}}$ (0.6-1.0; lower right panel). Low to high values are displayed in blue to red colour coding and the values for the parameters not varied are M$_\mathrm{He}$=4.0 M$_\odot$, E=1.0$\times$10$^{51}$ erg, M$_\mathrm{Ni}$=0.1 M$_\odot$ and Mix$_\mathrm{Ni}$=1.0.}
\label{f_mg_v_phot}
\end{figure}

Figure~\ref{f_mg_L} shows the dependence of the bolometric lightcurve on M$_{\mathrm{He}}$, E, M$_{\mathrm{Ni}}$ and Mix$_{\mathrm{Ni}}$, varying the parameters for a reference model with M$_{\mathrm{He}}$=4.0 M$_\odot$, E=1.0$\times$10$^{51}$ erg, M$_{\mathrm{Ni}}$=0.1 M$_\odot$ and Mix$_{\mathrm{Ni}}$=1.0. Qualitatively, we expect either an increase of the explosion energy or a decrease of the ejecta mass to decrease the diffusion time for thermal radiation, to decrease the optical depth for the $\gamma$-rays emitted in the decay chain of \element[ ][56]{Ni}, and to increase the expansion velocities. We therefore expect such a change to decrease the time at which peak luminosity occurs, to decrease the luminosity on the tail and to decrease the photospheric velocity. Qualitatively, we also expect the luminosity to scale with the mass of \element[ ][56]{Ni}. As seen in Fig.~\ref{f_mg_L}, all these qualitative dependencies are well followed by the models. Quantifying the dependencies by measuring the time (t$_{\mathrm{m}}$) and photospheric velocity (v$_{\mathrm{m}}$) at maximum luminosity (L$_{\mathrm{m}}$) for Mix$_{\mathrm{Ni}}$=1.0, and fitting a power-law expression to the model grid, we get

\begin{equation}
\mathrm{log~t_\mathrm{m} = 1.13 - 0.35~log~E_\mathrm{ej} + 0.58~log~M_\mathrm{ej} + 0.08~log~M_\mathrm{Ni}}
\label{eq_dep_1}
\end{equation}

\begin{equation}
\mathrm{log~v_\mathrm{m} = 1.09 + 0.43~log~E_\mathrm{ej} - 0.16~log~M_\mathrm{ej}}
\label{eq_dep_2}
\end{equation}

\begin{equation}
\mathrm{log~L_\mathrm{m} = 1.34 + 0.20~log~E_\mathrm{ej} - 1.02~log~M_\mathrm{ej} + 0.88~log~M_\mathrm{Ni}}
\label{eq_dep_3}
\end{equation}

which gives the (average) dependence of the observed quantities on the progenitor and SN parameters, expressed here for comparison in terms of the mass (M$_\mathrm{ej}$) and energy (E$_\mathrm{ej}$) of the ejecta. These are related to the helium core mass and explosion energy as $\mathrm{M_{ej}=M_{He}-M_{R}}$, where M$_\mathrm{R}$ is the mass of the compact remnant (1.5 M$_\odot$; see Sect.~\ref{s_sn_models}), and $\mathrm{E_{ej}=E+E_{\star}}$, where E$_\mathrm{\star}$ is the total (gravitational plus thermal) energy of the progenitor model (see Table \ref{t_mg_mesa_model}). Fitting the inverse relations\footnote{Note that the observed quantities are not necessarily independent.}
we get

\begin{equation}
\mathrm{log~E_\mathrm{ej} = -3.95 + 0.75~log~t_\mathrm{m} - 0.07~log~L_\mathrm{m} + 2.90~log~v_\mathrm{m}}
\label{eq_dep_4}
\end{equation}

\begin{equation}
\mathrm{log~M_\mathrm{ej} = -3.42 + 1.81~log~t_\mathrm{m} - 0.18~log~L_\mathrm{m} + 1.47~log~v_\mathrm{m}}
\label{eq_dep_5}
\end{equation}

\begin{equation}
\mathrm{log~M_\mathrm{Ni} = -4.96 + 2.08~log~t_\mathrm{m} + 0.93~log~L_\mathrm{m} + 1.19~log~v_\mathrm{m}}
\label{eq_dep_6}
\end{equation}

which gives the (average) dependence of the SN and progenitor parameters on the observed quantities. Using the observed values for SN 2011dh we get values for E$_{\mathrm{ej}}$, M$_{\mathrm{ej}}$ and M$_{\mathrm{Ni}}$ within $\sim$30 percent of those derived by the fitting procedure in Sect.~\ref{s_modelling_IIb}, and a clever parametrization of 
the observed quantities 
could actually be an alternative to this fitting procedure. However, given the 
doubtful 
results obtained for Type IIP SNe from the model grid fits by \citet{Lit83,Lit85} in e.g. \citet{Ham03}, care has to be taken, and we do not investigate this approach further in this work. The approximate model by \citet{Arn82} is often used to infer the SN and progenitor parameters for stripped envelope SNe \citep[e.g.][]{Lym14}. In this model the diffusion time and the expansion velocity depend on the mass and energy of the ejecta as $\mathrm{t_{d} \propto ({M_{ej}}^{3}/E_{ej})^{1/4}}$ and $\mathrm{v \propto (E_{ej}/M_{ej})^{1/2}}$, and inverting these gives $\mathrm{E_{ej} \propto {t_{d}}^2 v^3}$ and $\mathrm{M_{ej} \propto {t_{d}}^2 v}$. Comparing to the model grid fits we see that these scalings are qualitatively followed, but significant quantitative differences exist, and e.g.~E$_{\mathrm{ej}}$ is considerably less sensitive to t$_{\mathrm{m}}$. Clearly, t$_{\mathrm{m}}$ and t$_{\mathrm{d}}$ and in particular v and v$_{\mathrm{m}}$ are not identical, but most important is likely the fact that the \citet{Arn82} model assumes a constant opacity, whereas in the hydrodynamical models the (average) opacity is decreasing with time as the helium recombination front recedes through the ejecta (Sect.~\ref{s_physics}). An important consequence of the differences in the scalings is that t$_{\mathrm{m}}$ depends (roughly) on the quantity $\mathrm{{M_{ej}}^{2}/E_{ej}}$, whereas t$_{\mathrm{d}}$ depends on of the quantity $\mathrm{{M_{ej}}^{3}/E_{ej}}$ in the \citet{Arn82} model. This has implications for the degeneracy of the solution, as we will discuss further in Sect.~\ref{s_degeneracy}.

\subsection{Model physics}
\label{s_physics}

\begin{figure}[tbp!]
\includegraphics[width=0.5\textwidth,angle=0]{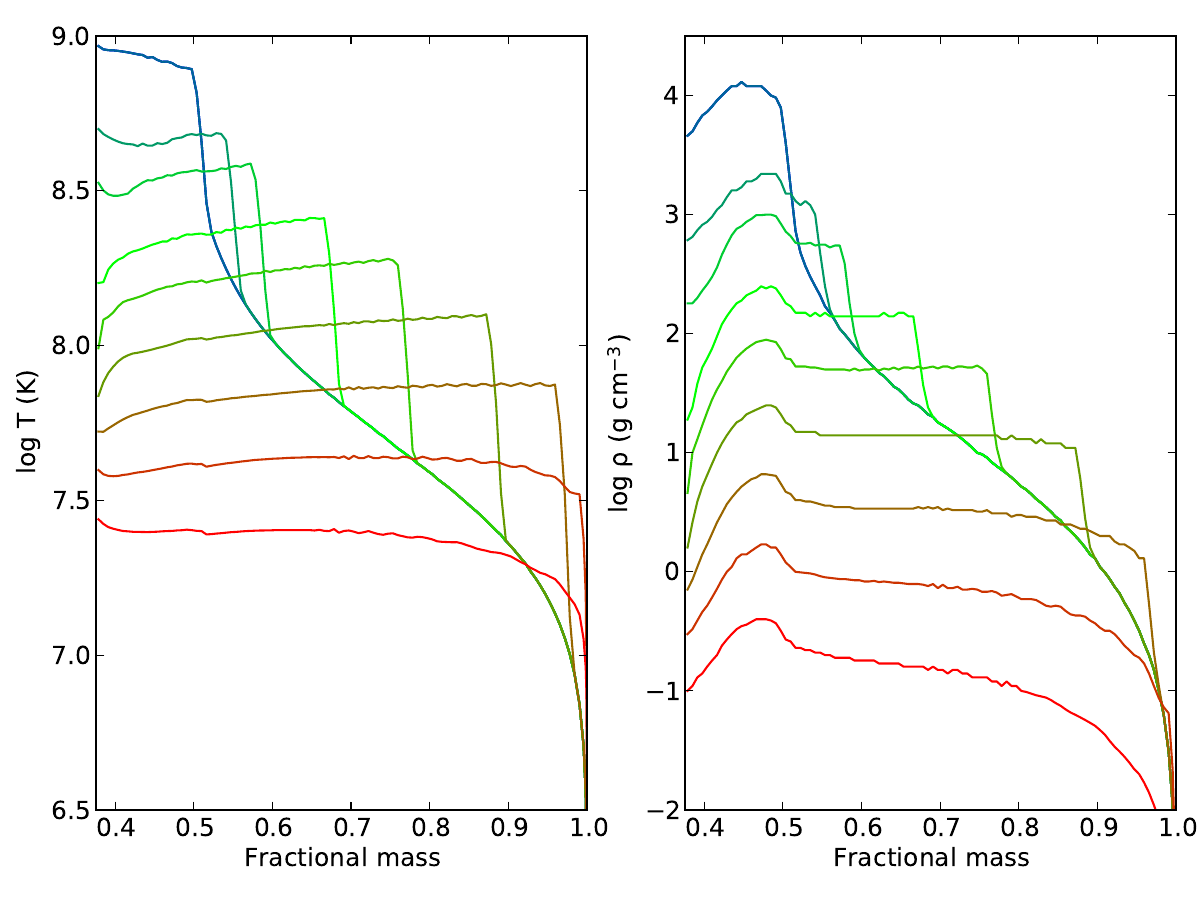}
\caption{Evolution of the temperature (left panel) and density (right panel) profiles between 1 and 282 seconds (shock breakout) in 10 logarithmically spaced intervals for the 4 M$_\odot$ helium-core model, where the time has been colour coded from blue (early) to red (late).}
\label{f_opt_model_rho_T_early}
\end{figure}

\begin{figure}[tbp!]
\includegraphics[width=0.5\textwidth,angle=0]{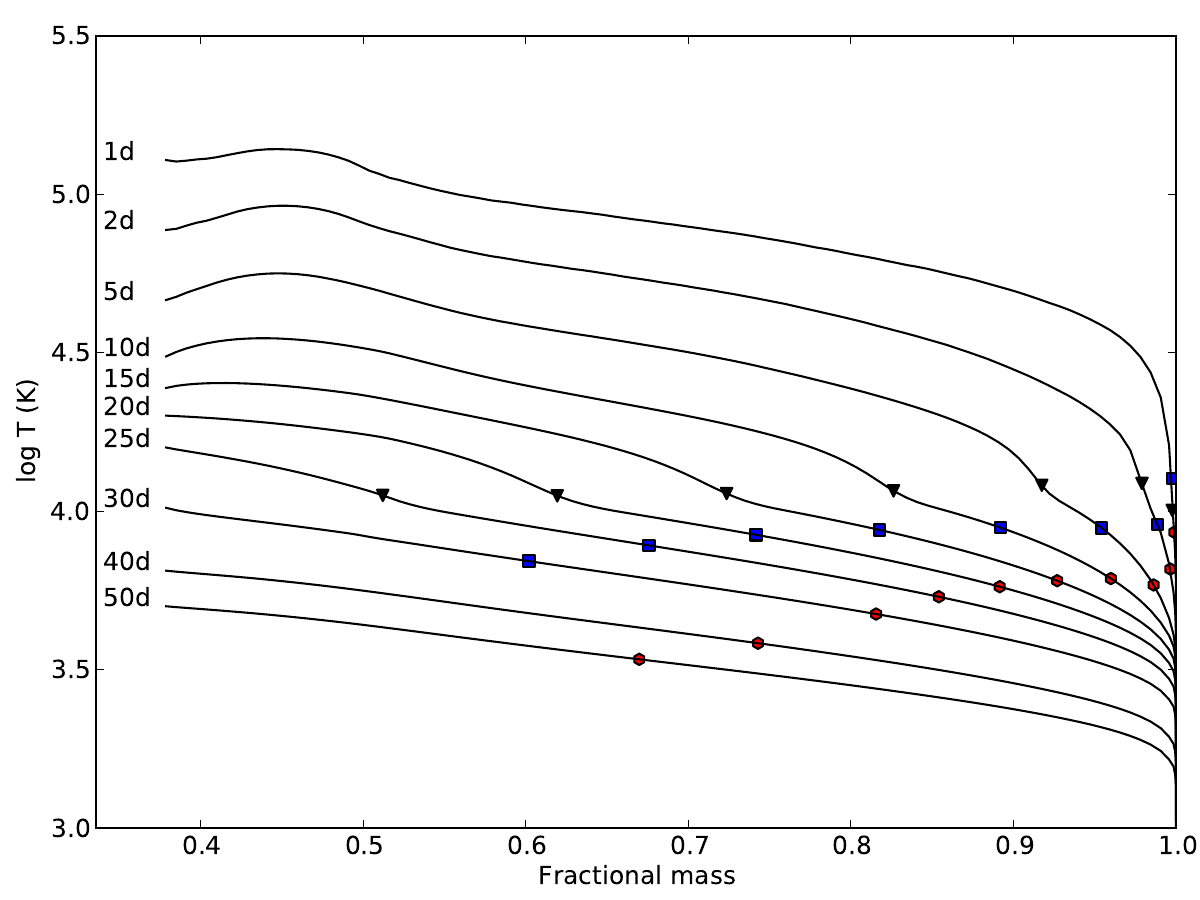}
\caption{Evolution of temperature profile for the 4 M$_\odot$ helium-core model. The position of the recombination front of helium (black triangles), the photosphere (red circles) and the thermalization surface (blue squares) have been marked, and each temperature profile annotated with the time since explosion.}
\label{f_opt_model_T}
\end{figure}

\begin{figure}[tbp!]
\includegraphics[width=0.5\textwidth,angle=0]{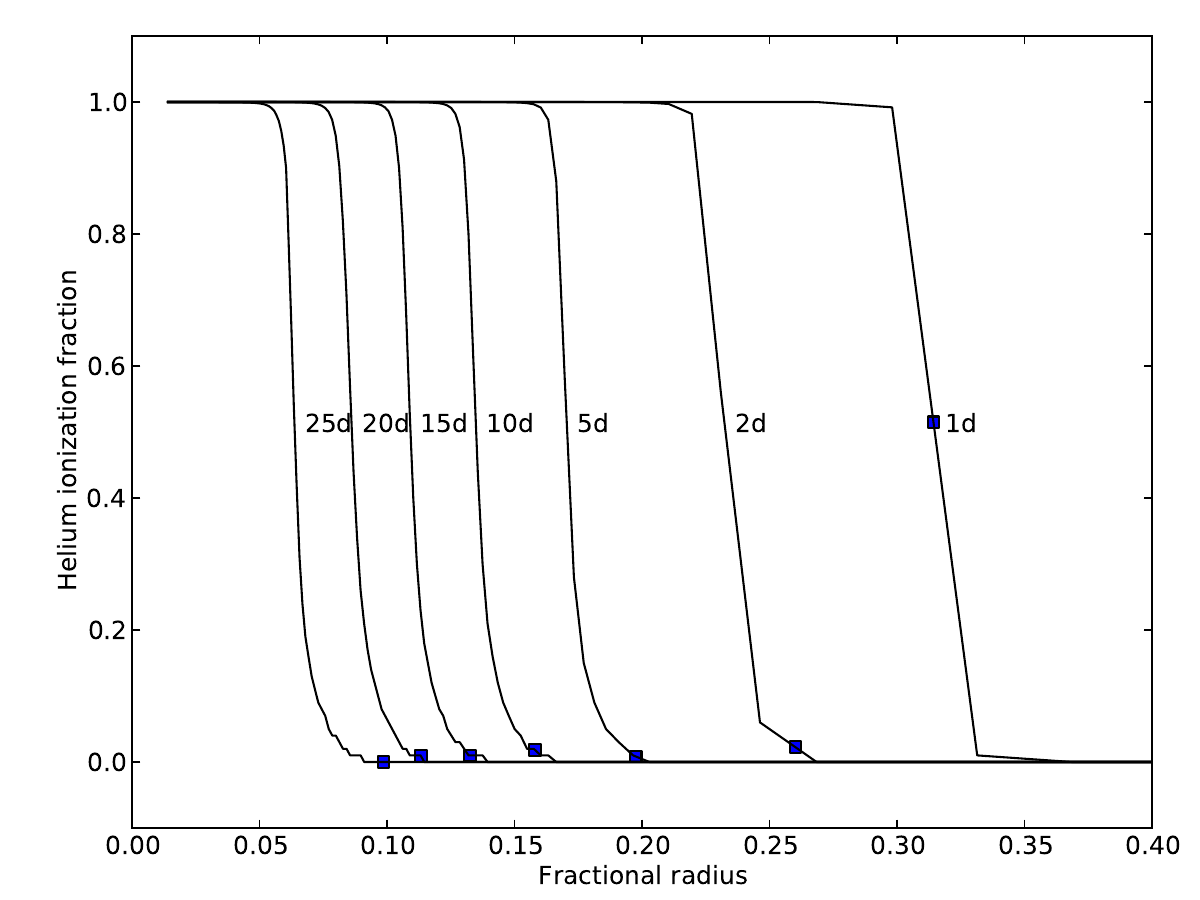}
\caption{Evolution of the helium ionization profile the 4 M$_\odot$ helium-core model. The thermalization surface (blue squares) have been marked and each ionization profile annotated with the time since explosion.}
\label{f_opt_model_He_ion}
\end{figure}

Here we discuss the physics of our bare helium-core models, exemplified by a 4 M$_\odot$ model with SN parameters E=1.0$\times$10$^{51}$erg, M$_{\mathrm{Ni}}$=0.1 M$_\odot$ and Mix$_{\mathrm{Ni}}$=1.0. We stress that the early evolution differs from that of an extended progenitor, which is discussed in Sect.~\ref{s_hydrogen_physics}.

Figure~\ref{f_opt_model_rho_T_early} shows the evolution of the density and temperature profiles from the injection of explosion energy until shock breakout, which occurs at $\sim$300 seconds. The shock initially accelerates to a speed of $\sim$10000 km~s$^{-1}$ in the oxygen core, but decelerates to $\sim$6000 km~s$^{-1}$ in the helium envelope, where the density gradient is small. In the outermost layers the density gradient increases drastically before it levels out in the thin convective envelope, and the shock accelerates to a speed of $\sim$30000 km~s$^{-1}$ at shock breakout. The thermal and kinetic energy behind the shock is close to equipartition and the temperature high enough for the equation of state to be completely radiation dominated. During the passage of the shock through the star some thermal energy is lost due to expansion, in particular during the passage through the thin envelope, and when the radiation breaks out from the shock the thermal fraction of the energy is $\sim$15 percent.

In the few minutes that follows the ejecta expand and the temperature and luminosity at the photosphere decrease rapidly by diffusion and expansion cooling. At $\sim$500 seconds the outermost parts become optically thin and the photosphere starts to recede into the ejecta. At $\sim$3 hours helium starts to recombine and at $\sim$10 hours the recombination front overruns the photosphere. Subsequently the position of the photosphere is determined by the recombination front, slowly moving inwards in mass coordinates, but outwards in radial coordinates. Figures~\ref{f_opt_model_T} and \ref{f_opt_model_He_ion} show the evolution of the temperature and the helium ionization profile between 1 and 50 and 1 and 25 days, respectively, where we have also marked the positions of the photosphere, the thermalization surface and the recombination front. The thermalization surface, here defined as $\sqrt {3 \tau_{\mathrm{abs}} \tau_{\mathrm{tot}}}=2/3$ \citep{Ens92}, is located near the outer edge of the recombination front, and follows the evolution of this until $\sim$25 days when the helium has recombined. During this period the temperature at the thermalization surface is roughly constant, and declines only slowly from $\sim$9000 K to $\sim$8000 K, a few thousand degrees below the temperature at the centre of the recombination front. 

We note that this temperature is in good agreement with the blackbody temperature measured for SN 2011dh in \citetalias{Erg14a}. \citet{Pir14} argue that the blackbody temperature for SN 2011dh and other similar SNe is too low to ionize helium and that their helium envelopes might be effectively transparent, but according to our results this is not the case. Instead, we find that the gradual recombination of the helium envelope is actually what shapes the major part of the diffusion peak lightcurve for a Type IIb SN.

\subsection{Model limitations}
\label{s_model_limitations}

\begin{figure}[tbp!]
\includegraphics[width=0.5\textwidth,angle=0]{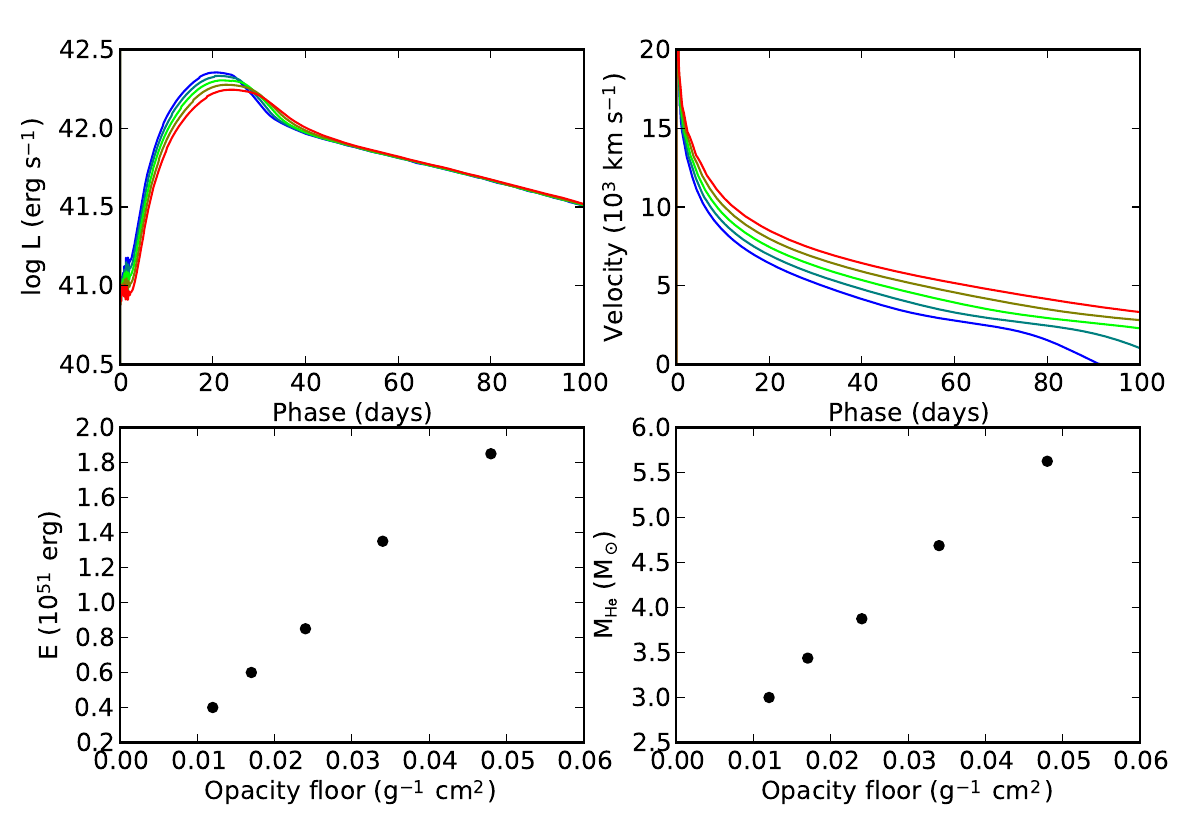}
\caption{Upper panels: Model bolometric lightcurves (left panel) and photosheric velocities (right panel) for day 1-100 showing the dependence on the opacity floor (0.012-0.048 g$^{-1}$ cm$^{2}$). Low to high values are displayed in blue to red colour coding and the model parameters are M$_{\mathrm{He}}$=4.0 M$_\odot$, E=1.0$\times$10$^{51}$ erg, M$_{\mathrm{Ni}}$=0.1 M$_\odot$ and Mix$_{\mathrm{Ni}}$=1.0. Lower panels: Sensitivity of the explosion energy (left panel) and helium-core mass (right panel) to the opacity floor, calculated by fitting the model lightcurves and photospheric velocities to the model grid using the procedure described in Sect.~\ref{s_fitting_proc}.}
\label{f_kappa_floor}
\end{figure}

Although more sophisticated than the \citet{Arn82} like models often used in SN sample studies, the hydrodynamical models used in this study still suffer from a number of limitations. Below we discuss briefly the most important of these and the limitations in our treatment of the opacity in some more detail, as the bolometric lightcurves depend critically on this quantity.

\paragraph{Progenitor models} 

The progenitor models only differ in helium-core mass, and are all non-rotating and with solar metallicity. This is clearly a simplification, but as the number of SN models would increase drastically we have chosen not to vary these progenitor parameters. The effect of a low-mass hydrogen-envelope, not present in our bare helium-core models, is discussed separately in Sect.~\ref{s_hydrogen_envelope}.

\paragraph{Hydrodynamics} 

As HYDE does not include a nuclear reaction network the effect of the explosive nucleosynthesis in the inner part of the ejecta is not included, and as HYDE is 1-D, macroscopic mixing of the nuclear burning zones (see \citealp{Iwa97}) is prohibited. The mass-cut is fixed at 1.5 M$_\odot$ and although fallback of material onto the compact remnant is not prohibited, the artificial zero-velocity inner boundary condition will cause this material to bounce. To properly handle fallback a piston-driven explosion without an inner boundary condition on the velocity would be needed. 

\paragraph{Radiative transfer} 

The optically thin regime is not handled correctly as the code is based on the 
diffusion approximation. 
The treatment of the optically thin region is critical when calculating spectra or broad-band photometry, but probably of less importance when calculating the bolometric lightcurve. A correct treatment of the optically thin region could also be important for the radiative acceleration of the outer parts of the ejecta occurring at early times. This could effect the bolometric lightcurve during the cooling phase, but probably not later on.

\paragraph{Opacity}

One major limitation with HYDE is the absence of a proper treatment of the line (bound-bound) opacity. The opacity tables used are calculated for a static medium and does not take into account the effect of a velocity gradient, which tend to increase the line opacity \citep[e.g.][]{Kar77}. Furthermore, as the opacity is calculated for a medium in LTE, it may not apply in the optically thin region, where non-thermal ionization could increase the electron scattering contribution. To compensate for this lack of opacity, 
HYDE makes use of an opacity floor (Sect.~\ref{s_hyde_opacity}). The value of this floor is set to 0.024 g$^{-1}$ cm$^{2}$, which is much lower than the electron scattering opacity of $\sim$0.2 g$^{-1}$ cm$^{2}$ for fully ionized helium-core material, and only affects the region outside the recombination front (Sect.~\ref{s_physics}). The upper panels of Fig.~\ref{f_kappa_floor} show the dependence of the model lightcurves and photospheric velocities on the opacity floor for a 4 M$_\odot$ model with explosion parameters E=1.0$\times$10$^{51}$erg, M$_{\mathrm{Ni}}$=0.1 M$_\odot$ and Mix$_{\mathrm{Ni}}$=1.0. These are not particularly sensitive, and a doubling of the opacity floor corresponds to an increase in the photospheric velocities and a shift to later times of the diffusion peak of 10-15 percent. The lower panels of Fig.~\ref{f_kappa_floor} show the sensitivity of the estimated helium-core mass and explosion energy to the opacity floor, calculated by fitting the model lightcurves and photospheric velocities to the model grid using the same procedure as in Sect.~\ref{s_modelling_IIb}. The explosion energy depends strongly on the photospheric velocities (Eqs.~\ref{eq_dep_4}), so this quantity is rather sensitive and almost proportional to the opacity floor, whereas the helium-core mass is less affected and a doubling of the opacity floor corresponds to an increase of $\sim$25 percent. A proper investigation of the effects of the opacity floor on our results in Sect.~\ref{s_modelling_IIb} would require a 
comparison with a code capable of calculating the line opacity correctly, and is outside the scope of this paper. We conclude, however, that the helium-core mass is not particularly sensitive to the choice of opacity floor, whereas this choice is more critical with respect to the explosion energy.

\section{The hydrogen envelope}
\label{s_hydrogen_envelope}

As our aim is to use the grid of bare helium-core models to fit the bolometric lightcurves and the photospheric velocities of Type IIb SNe, which may have extended low-mass hydrogen envelopes surrounding their helium cores, it is of importance to investigate which effect such envelopes would have on the observed properties, as well as on the results obtained in Sect.~\ref{s_modelling_IIb}. It is also of interest to discuss the physics of such models and compare to the physics of bare helium (Sect.~\ref{s_physics}).

\subsection{Effect on the observed properties}
\label{s_hydrogen_effects}

\begin{figure}[tb!]
\includegraphics[width=0.5\textwidth,angle=0]{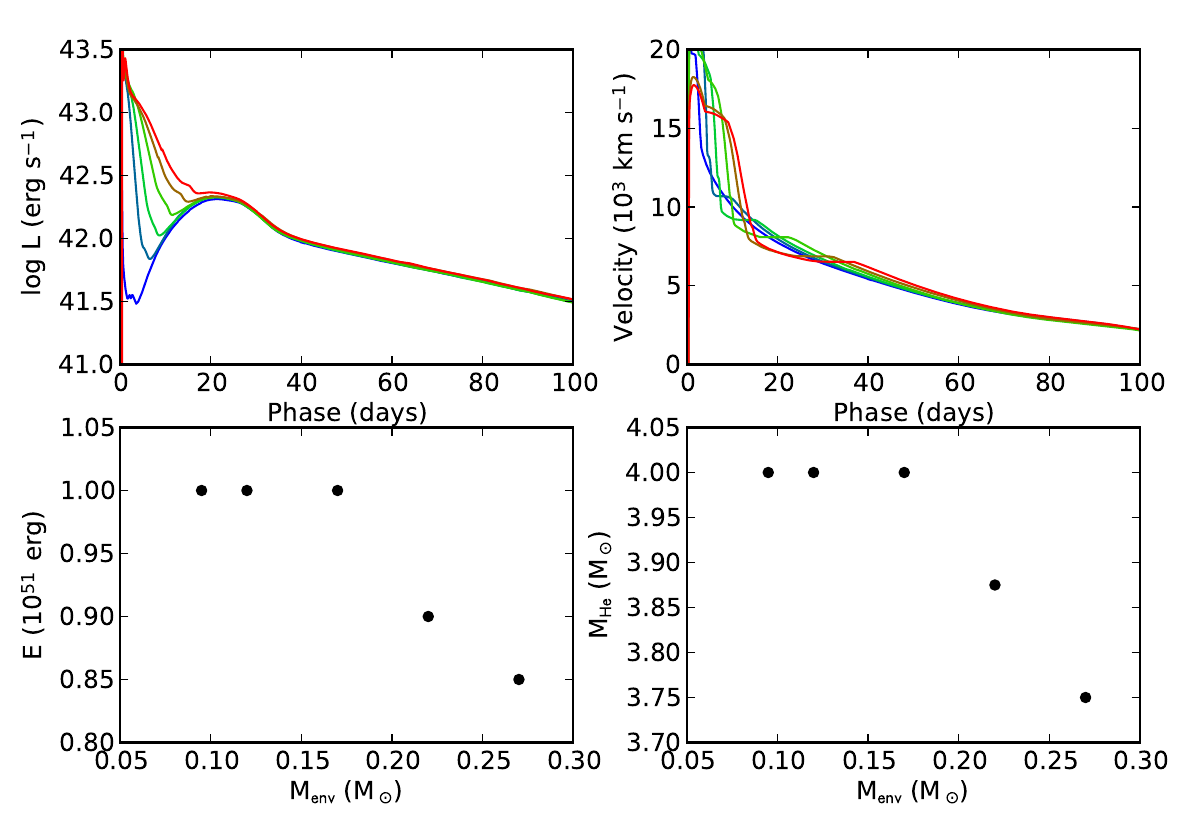}
\caption{Upper panels: Progression of model bolometric lightcurves (left panel) and photospheric velocities (right panel) calculated with HYDE for a 15 M$_\odot$ MESA model with the mass-loss adjusted to yield a final mass of 4.2, 4.15, 4.1, 4.05, 4.025 and 4.0 M$_\odot$, colour coded from red (4.2 M$_\odot$) to blue (4.0 M$_\odot$). The SN parameters were E=1.0$\times$10$^{51}$erg, M$_{\mathrm{Ni}}$=0.1 M$_\odot$ and Mix$_{\mathrm{Ni}}$=M$_{\mathrm{He}}$/M. Lower panels: Sensitivity of the explosion energy (left panel) and helium-core mass (right panel) to the mass of the hydrogen envelope, calculated by fitting the model lightcurves and photospheric velocities to the model grid using the procedure described in Sect.~\ref{s_fitting_proc}.}
\label{f_hydrogen}
\end{figure}

The upper panels of Fig.~\ref{f_hydrogen} show the bolometric lightcurve and photospheric velocities for a sequence of 15 M$_\odot$ MESA models, where the mass loss was adjusted to yield final masses in the range 4.2-4.0 M$_\odot$. Defining the hydrogen envelope to begin where X>0.01, this corresponds to hydrogen envelope masses in the range 0.27-0.07 M$_\odot$. The bolometric lightcurves for all models show an initial decline phase corresponding to the cooling of the thermal explosion energy deposited in the hydrogen envelope, the length of which decreases with 
the 
mass of the envelope. The reason for this is twofold, first the thermal energy deposited in the hydrogen envelope decreases with the mass of it, and secondly the radius of the progenitor stars decreases, which decreases the time scale for expansion cooling. Models for Type IIb SNe often have an increased helium abundance in the hydrogen envelope \citep[e.g.][]{Woo94}, due to mixing of helium into the base of the hydrogen envelope. This results in smaller progenitor radii due to decreased opacities, and therefore in shorter durations of the cooling phase. Our models have lower helium abundances as compared to \citet{Woo94} and \citet{Shi94}, which should be kept in mind. During most of the cooling phase the photospheric velocities are much higher than those for a bare helium-core model, but decrease quickly at the luminosity minimum, after which follows a period when they are significantly lower. The latter effect is larger for models with more massive hydrogen envelopes and is likely caused by deceleration of the helium core. As suggested by Eqs.~\ref{eq_dep_4}-\ref{eq_dep_5}, and as demonstrated in Sect.~\ref{s_error_sensitivity}, the sensitivity of the estimated explosion energy to an error in the photospheric velocity is high (E $\sim$ v$^{3}$), whereas the sensitivity of the helium-core mass is lower. This is quantified by the lower panels of Fig.~\ref{f_hydrogen}, which show the sensitivity of the estimated explosion energy and helium-core mass to the mass of the hydrogen envelope, calculated by fitting the model lightcurves and photospheric velocities to the model grid using the same procedure as in Sect.~\ref{s_modelling_IIb}. If the mass of the hydrogen envelope is larger than $\sim$0.2 M$_\odot$ the explosion energy is significantly underestimated, whereas the helium-core mass is less affected. Therefore the presence of a relatively massive hydrogen envelope could have a significant effect on the estimated explosion energy when using bare helium-core models. It is worth noting that if we subtract the energy in the hydrogen envelope, the 
ejecta 
energy is in much better agreement with what is estimated from the fit, which therefore rather provide an estimate of the energy in the helium core. 
Except for the effect on the photospheric velocities, the presence of the hydrogen envelope does not seem to appreciably affect the observed properties after the luminosity minimum.

\subsection{Model physics}
\label{s_hydrogen_physics}

Here we discuss the physics of models with low-mass hydrogen envelopes, exemplified by the 4.05 M$_\odot$ model shown in Fig.~\ref{f_hydrogen}. This model has an hydrogen envelope of 0.17 M$_\odot$, an average hydrogen fraction in the envelope of 0.5, and reach the luminosity minimum at $\sim$11 days, which is similar to, but slightly later than was observed for SN 1993J.

The passage of the shock through the helium core proceeds as described for the bare helium-core model, and in the steep density gradient between the helium core and the hydrogen envelope it accelerates to $\sim$20000 km~s$^{-1}$. Once in the hydrogen envelope, where the density is roughly constant, the shock gradually decelerates to $\sim$6000 km~s$^{-1}$, which gives rise to a reverse shock propagating backwards into the helium envelope. During the passage of the shock through the hydrogen envelope, the helium core expands and most of the deposited thermal energy is cooled away. At shock breakout the (relatively) cool and expanded helium core is surrounded by the hot and compressed hydrogen envelope, and the subsequent evolution is determined by the expansion and cooling of this envelope. The fraction of the energy deposited in the envelope is about 10 percent, roughly equipartioned into thermal and kinetic energy.

\begin{figure}[tbp!]
\includegraphics[width=0.5\textwidth,angle=0]{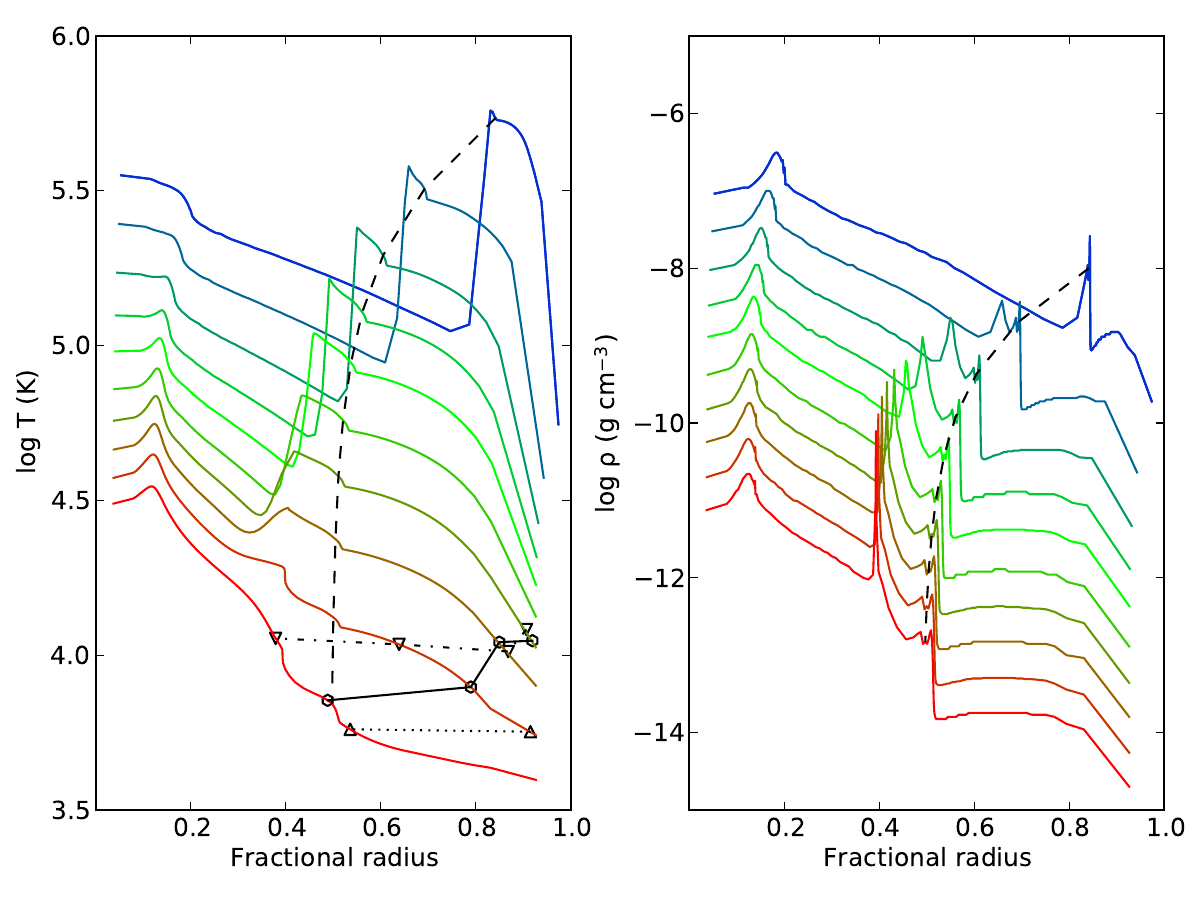}
\caption{Evolution of the temperature (left panel) and density (right panel) profiles between shock breakout (0.3 days) and the luminosity minimum (11 days) in 10 logarithmically spaced intervals for the 4.05 M$_\odot$ MESA model, where the time has been colour coded from blue (early) to red (late). The interface between the helium and hydrogen envelopes (dashed line), the photosphere (circles and dashed line) and the helium (downward triangles and dot-dashed line) and hydrogen (upward triangles and dotted line) recombination fronts are also shown.}
\label{f_hydrogen_rho_T_early}
\end{figure}

Figure~\ref{f_hydrogen_rho_T_early} shows the evolution of the density and temperature profiles from shock breakout, which occurs at $\sim$0.3 days, until the luminosity minimum. The profiles are similar to those obtained by modelling of SN 1993J in \citet{Woo94}, \citet{Shi94} and \citet{Bli98}. Initially the hydrogen envelope is opaque and ionized, and the surface luminosity and temperature decreasing by expansion cooling, but at $\sim$4 days the outer parts become optically thin and the photosphere starts to recede into the ejecta. The helium starts to recombine at about the same time, whereas hydrogen stays ionized until $\sim$7 days, and at about $\sim$6 days the photosphere starts to trace the helium recombination front as in the bare helium-core models. At the luminosity minimum the hydrogen in the envelope has recombined, the photosphere is located close to the interface between the hydrogen and helium envelope, and the temperature at the thermalization surface is $\sim$9500 K.

\section{Model grid fits}
\label{s_modelling_IIb}

Here we use an automated procedure to fit the bolometric lightcurves and photospheric velocities for the CSP and literature samples of Type IIb SNe to those of our grid of SN models\footnote{The results for SNe 2003bg, 1996cb, 2011ei, and 2011fu were obtained with an earlier version of the model grid, which is based on (mass) scaled versions of the 4 M$_\odot$ bare helium-core model from \citet{Nom88} and a fixed mixing of the $^{56}$Ni (M$_{\mathrm{Ni}}$=1.0)}. Our method allows us to determine the sensitivity of the derived quantities to errors in the observed quantities, as well as to investigate the degeneracy of the solutions found. As discussed, the grid is based on bare helium-core models, and given that a low-mass hydrogen envelope would mainly affect the cooling phase (Sect.~\ref{s_hydrogen_effects}), the diffusion phase and early tail lightcurves, and the diffusion phase photospheric velocities, are used to determine the helium core and SN parameters. However, keep in mind that relatively massive hydrogen envelopes do affect the photospheric velocities in the diffusion phase, which might lead to significant underestimates of the explosion energy (Sect.~\ref{s_hydrogen_effects}). We also make the assumption, justified for Type IIb SNe, that the helium core is not affected by mass loss. 

\subsection{The Type IIb SNe sample}

The Type IIb SNe in the CSP stripped-envelope SNe sample (Stritzinger~et~al.~2015) 
consist of SNe 2004ex, 2004ff, 2005Q, 2006T, 2006ba, 2006bf, 2008aq, 2009K, 2009Z and 2009dq, of which SNe 2006bf and 2009dq have been excluded due to bad sampling of the lightcurves. The Type IIb SNe that have been individually studied in the literature consist of SNe 1993J \citep[e.g.][]{Ric94,Ric96}, 1996cb \citep{Qiu99}, 2003bg \citep{Hamu09}, 2008ax \citep[e.g.][]{Tau11}, 2010as \citep{Fol14b}, 2011dh \citepalias[e.g.][]{Erg14a}, 2011fu \citep{Kum13}, 2011hs \citep{Buf14}, 2011ei \citep{Mil13} and 2013df \citep{Dyk14}, of which the data for SN 2010as was not yet available when this work began, and is therefore not included. Observations of 4 additional Type IIb SNe (2001gd 2006el 2008bo and 2008cw) have been published as part of surveys, but are not included in our sample. Out of the SNe in the sample, 1993J, 2008ax and 2011dh stands out by the quality of the data as well as the hard constraints on the explosion epochs. The 
observational details, 
the constraints on the explosion epochs and the adopted distances and extinctions for the CSP sample and the sample of individually studied SNe  
are given in Appendices~\ref{a_csp_sample} and \ref{a_literature_sample}, respectively, where we also describe how the photospheric velocities were estimated. The pseudo-bolometric lightcurves were calculated from the photometry using the methods described in \citetalias{Erg14a}, and a UV to MIR bolometric correction (BC) determined from SN 2011dh applied. The flux falling outside this wavelength range was not corrected for, but given the results from the steady-state NLTE modelling of SN 2011dh presented in \citet[][hereafter \citetalias{Erg14b}]{Erg14b}, this correction is likely to be small (<0.15 mag).

\subsection{Fitting procedure}
\label{s_fitting_proc}

The fitting is done by minimization of the square of the relative residuals, giving equal weight to the diffusion phase lightcurve, the tail lightcurve and the diffusion phase photospheric velocity evolution. The division between the diffusion and tail phases is made roughly at the point where the decline rate of the bolometric lightcurve becomes constant. If there is any sign of a cooling phase, the beginning of the diffusion phase is set to a few days after the rise to peak begins, and otherwise to the first observation. Photospheric velocities above the interface between the helium core and the hydrogen envelope are excluded from the fit. As discussed in \citetalias{Erg14a} this velocity can be estimated from the minimum velocity for the  H$\alpha$ absorption minimum, but is otherwise set to 10000 km~s$^{-1}$. As the explosion epochs in many cases are not well constrained we fit, not only the progenitor (M$_{\mathrm{He}}$) and SN (E, M$_{\mathrm{Ni}}$ and Mix$_{\mathrm{Ni}}$) parameters, but also the epoch of explosion, which is allowed to vary between the hard limits obtained from detections and non-detections. The errors in the bolometric lightcurves arising from the uncertainties in distance and extinction, and a systematic error in the photospheric velocities, assumed to be 15 percent, were propagated by standard methods.

\subsection{Results and comparisons}

\begin{figure*}[tb!]
\includegraphics[width=1.0\textwidth,angle=0]{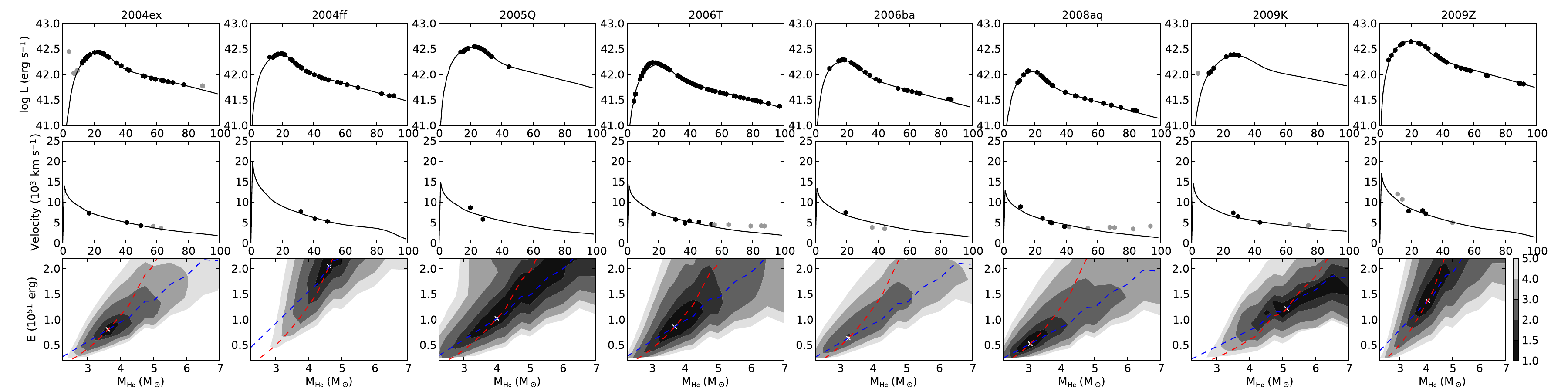}
\caption{Bolometric lightcurve (upper panels) and photospheric velocity evolution (middle panels) for the best-fit models as compared to the observed UV to MIR pseudo-bolometric lightcurve and estimated photospheric velocity evolution for the CSP sample of Type IIb SNe. The lower panels show contour plots of the standard deviation in the fits, normalized to that of the optimal model, projected onto the E-M$_{\mathrm{He}}$ plane. We also show the constraints $\mathrm{M_{ej}/E_{ej}=\text{const}}$ (blue) and $\mathrm{{M_{ej}}^{2}/E_{ej}=\text{const}}$ (red) provided by the photospheric velocity evolution and the bolometric lightcurve, respectively.}
\label{f_mg_fit_csp_comp}
\end{figure*}

\begin{figure*}[tb!]
\includegraphics[width=1.0\textwidth,angle=0]{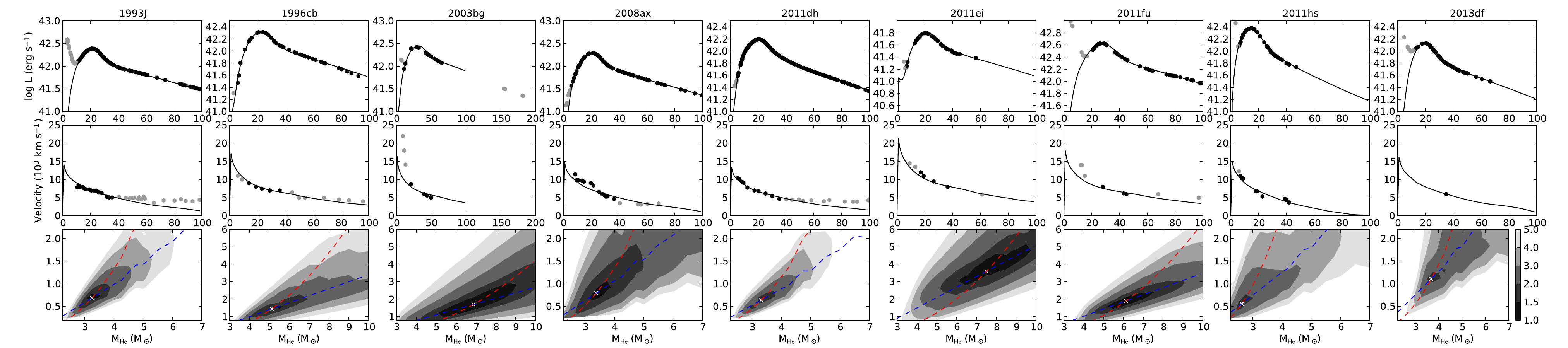}
\caption{Bolometric lightcurve (upper panels) and photospheric velocity evolution (middle panels) for the best-fit models as compared to the observed UV to MIR pseudo-bolometric lightcurve and estimated photospheric velocity evolution for the sample of individually studied Type IIb SNe. The lower panels show contour plots of the standard deviation in the fits, normalized to that of the optimal model, projected onto the E-M$_{\mathrm{He}}$ plane. We also show the constraints $\mathrm{M_{ej}/E_{ej}=\text{const}}$ (blue) and $\mathrm{{M_{ej}}^{2}/E_{ej}=\text{const}}$ (red) provided by the photospheric velocity evolution and the bolometric lightcurve, respectively.}
\label{f_mg_fit_lit_comp}
\end{figure*}

Figures~\ref{f_mg_fit_csp_comp} and \ref{f_mg_fit_lit_comp} show the best-fit model bolometric lightcurve and photospheric velocity evolution, compared to the observed UV to MIR pseudo-bolometric lightcurve and estimated photospheric velocity evolution, as well as contour plots of the standard deviation in the fits, normalized to that of the optimal model, projected onto the E-M$_{\mathrm{He}}$ plane. Tables~\ref{t_mg_fit_csp_comp} and \ref{t_mg_fit_lit_comp} give the helium-core mass, explosion energy, mass and mixing of \element[ ][56]{Ni}, and explosion epoch for the best-fit models and the corresponding errors\footnote{The zero error bars given in some case arise by technical reasons and will be fixed in the final version of the paper.}. The fits are mostly good, and for SNe 1993J, 2004ex, 2004ff, 2006T, 2006ba, 2008aq, 2008ax, 2011dh, 2011hs and 2013df the solutions are well constrained in the the E-M$_{\mathrm{He}}$ plane, although SN 2011hs is not well constrained below as it lies at the border of the covered parameter space. For SNe 2003bg, 2009T, 2009K and 2011ei, the constraint from the bolometric lightcurve is weak due to the limited coverage, and the solutions are quite degenerate along the M/E=const curve (see Sect.~\ref{s_degeneracy}). A significant degeneracy along the M/E=const curve is also seen for SNe 1996cb and 2011fu, although the lightcurve coverage for these SNe is much better, whereas for SN 2009Z, the solution is a bit degenerate along the M$^{2}$/E=const curve (which would suggest a weak constraint from the velocity, see Sect.~\ref{s_degeneracy}). 

\begin{figure}[tb!]
\includegraphics[width=0.48\textwidth,angle=0]{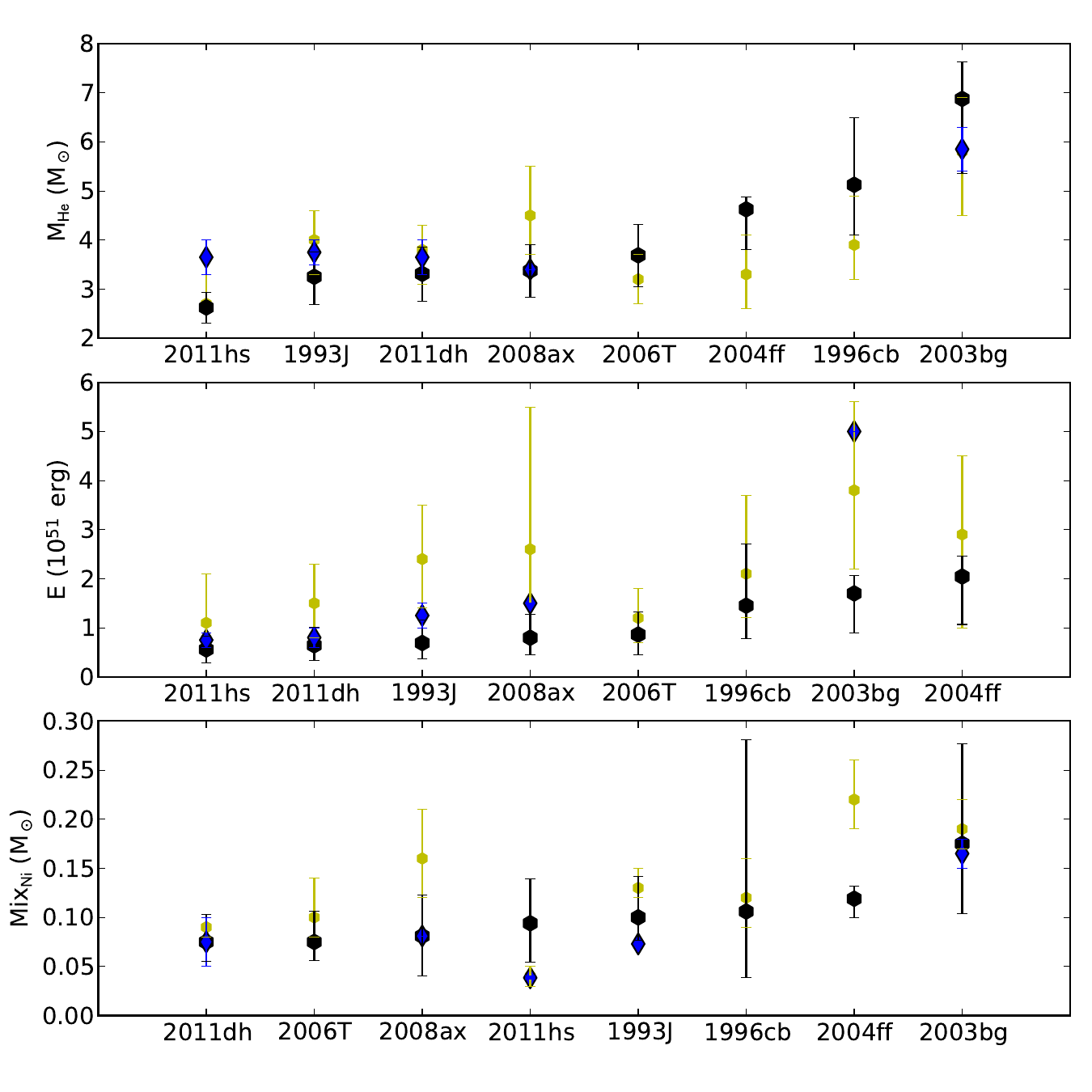}
\caption{Comparison of the helium-core mass (upper panel), explosion energy (middle panel) and mass of \element[ ][56]{Ni} as estimated using hydrodynamical modelling in this (black circles) and other (blue diamonds) works, and as estimated using approximate \citet{Arn82} models by \citet{Lym14} (yellow circles), where in each panel the SNe have been ordered with respect to the values estimated in this work. In cases where the results are given in terms of the mass and energy of the ejecta, a mass of 1.5 M$_\odot$ has been assumed for the compact remnant, and the difference between the explosion and ejecta energy ignored.}
\label{f_literature_compare}
\end{figure}

Several SNe (1993J, 2003bg, 2008ax, 2011dh and 2011hs) in our sample have been studied using hydrodynamical modelling in other works, and eight of the SNe are also included in the sample study by \citet{Lym14}, based on approximate \citet{Arn82} models. Figure~\ref{f_literature_compare} shows as comparison of the helium-core mass, explosion energy and mass of \element[ ][56]{Ni} estimated in these works with our results. In cases where the results are given in terms of the mass and energy of the ejecta, a mass of 1.5 M$_\odot$ has been assumed for the compact remnant, and the difference between the explosion and ejecta energy ignored. Turning first to the results obtained with hydrodynamical modelling, the agreement is mostly reasonable, but some clear differences exist. In particular the explosion energies estimated in our study for SNe 1993J, 2003bg and 2008ax are considerable lower than in other works. SN 1993J has been modelled by \citet{Woo94} and \citet{Shi94} among others, SN 2003bg by \citet{Maz09} and SN 2008ax by \citet{Tsv09}. In the case of SNe 1993J and 2003bg the disagreement could be caused by the presence of a relatively massive hydrogen envelope (Sect.~\ref{s_hydrogen_envelope}), as suggested by the extent of the cooling phase in the case of SN 1993J. As discussed in Sect.~\ref{s_hydrogen_effects}, the explosion energy estimated from the fit is rather an estimate of the explosion energy deposited in the helium core. According to \citet{Shi94} and \citet{Maz09}, this energy was $\sim$0.6$\times$10$^{51}$ and $\lesssim$2.5$\times$10$^{51}$ erg in their models of SNe 1993J and 2003bg, respectively, which is in better agreement with our results, and in support of the hypothesis. However, in this case of SN 2008ax, the situation could be the reverse. The absence of an extended cooling phase suggests a relatively low-mass hydrogen envelope, so the SN 1993J based model used by \citet{Tsv09} could lead to a significant overestimate of the explosion energy. 

For SN 2011hs, which was modelled by \citet{Buf14}, we find a significantly lower helium core mass, and a significantly higher mass of \element[ ][56]{Ni}. The difference in the mass of \element[ ][56]{Ni} can be traced back to differences in the adopted extinction and distance (Appendix~\ref{a_literature_sample}), and with respect to the helium-core mass it is worth noting that models with M$_\mathrm{He}$<3.3 M$_\odot$ was not tested by \citet{Buf14}, so their modelling do not provide a lower bound on this quantity. For SN 2011dh our results are in good agreement with those in \citet{Ber12}, and also with those obtained in \citetalias{Erg14b}, which are based on the <400 days lightcurve and an extended version of the model grid using a BC determined with steady-state NLTE modelling. They are also further supported by the modelling of nebular spectra in \citet{Jer14} and the stellar evolutionary progenitor analysis by \citet{Mau11}.

Comparing to the results from the sample study by \citet{Lym14} we see a reasonable agreement with respect to the helium-core masses and the masses of \element[ ][56]{Ni}, although the latter differ considerably in a few cases, 
a fact that 
can be traced back to the adopted extinctions and distances (Appendices~\ref{a_csp_sample} and \ref{a_literature_sample}). On the other hand, the explosion energies estimated by \citet{Lym14} are systematically higher as compared to our results. As the \citep{Arn82} model is considerably simpler than our hydrodynamical models, the explanation cannot be the same as discussed above, and is rather related to the unclear relation between expansion velocity and the observed velocities in the \citet{Arn82} model. This results in a considerable uncertainty, propagating mainly to the estimated explosion energy (see Sect.~\ref{s_error_sensitivity} with respect to the hydrodynamical models). Nevertheless, the explosion energy estimated for SN 2003bg by \citet{Lym14} is actually in better agreement with the hydrodynamical modelling by \citet{Maz09} than our results. We speculate that the reason for this is that \citet{Lym14} include the high photospheric velocities in the hydrogen envelope, excluded for consistency in our bare helium-core fits (Sect.~\ref{s_fitting_proc}), and therefore arrives at an estimate of the total kinetic energy in the ejecta. Further comparisons with the results by \citet{Lym14} are made in Sect.~\ref{s_sample_statistics}, where we discuss the sample statistics. Finally, it is worth noting that the small ejecta mass of 0.3 M$_\odot$ estimated for SN 2011ei from modelling of nebular spectra in \citet{Mil13} seems to be excluded by our results. This SN is quite interesting, as it shows the most extreme values for the progenitor and SN parameters in the sample. Whereas the helium-core mass and explosion energy are the highest, the mass of \element[ ][56]{Ni} is the lowest. The high helium-core mass and explosion energy derived stem mainly from the unusually high photospheric velocities, whereas the time at which peak luminosity occurs is quite typical. 

\begin{table*}[tb!]
\caption{Explosion energy, helium-core mass, mass and mixing of the \element[ ][56]{Ni} and epoch of explosion for the best-fit models for the CSP sample of Type IIb SNe.}
\begin{center}
\include{mg-fit-csp-comp-table}
\end{center}
\label{t_mg_fit_csp_comp}
\end{table*}

\begin{table*}[tb!]
\caption{Explosion energy, helium-core mass, mass and mixing of the \element[ ][56]{Ni} and epoch of explosion for the best-fit models for the sample of individually studied Type IIb SNe.}
\begin{center}
\include{mg-fit-lit-comp-table}
\end{center}
\label{t_mg_fit_lit_comp}
\end{table*}

\subsection{Error sensitivity}
\label{s_error_sensitivity}

\begin{figure}[tb]
\includegraphics[width=0.5\textwidth,angle=0]{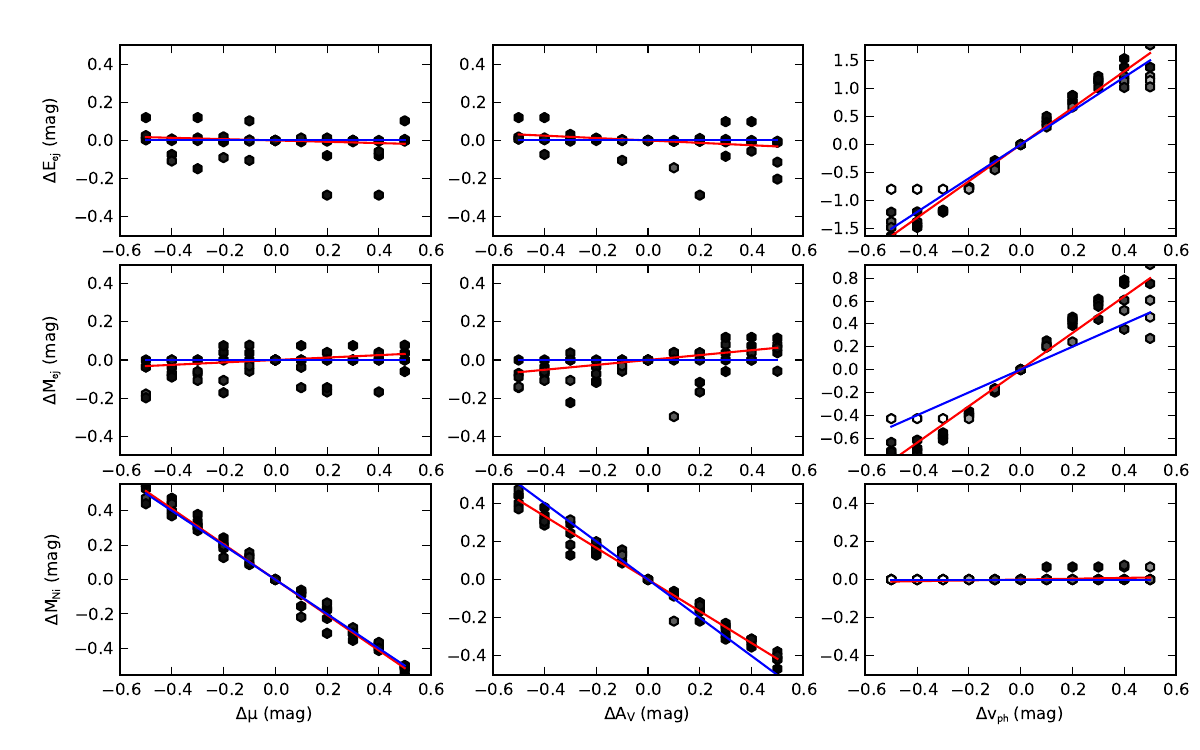}
\caption{Sensitivity of the derived ejecta energy (upper panels), ejecta mass (middle panels) and mass of \element[ ][56]{Ni} (lower panels) to a change in the distance (left panels), extinction (middle panels) and photospheric velocities (right panels). For consistency the changes in all quantities are expressed in magnitudes. The derived quantities for our sample of Type IIb SNe are shown as black dots, power-law fits as red solid lines and the scalings expected from the \citet{Arn82} model as blue solid lines.}
\label{f_mg_error_prop}
\end{figure}

 Figure~\ref{f_mg_error_prop} shows the sensitivity of the derived quantities to errors in the distance, extinction and a systematic error in the photospheric velocity for our sample of Type IIb SNe\footnote{The current version of the figure only include SNe 1993J, 2008ax, 2011dh, 2011hs and 2013df.}. To compare with approximate scalings the derived quantities are expressed in terms of the mass and energy of the ejecta, but this does not affect the conclusions. For the helium-core mass and explosion energy the dependence on the distance and extinction is weak, whereas the dependence on the photospheric velocity is strong. For the mass of \element[ ][56]{Ni} the sensitivity on the distance and extinction is strong, whereas the dependence on the photospheric velocity is weak. In general we see that an error in the distance and extinction mainly corresponds to an error in the mass of \element[ ][56]{Ni}, whereas an error in the photospheric velocity mainly corresponds to an error in the helium-core mass and explosion energy. This behaviour is in agreement with the model grid dependencies discussed in Sect.~\ref{s_dependence} and with the qualitative discussion in \citetalias{Erg14a}, based on the \citet{Arn82} model. In Fig.~\ref{f_mg_error_prop} we show the scalings expected from the \citet{Arn82} model and, as previously discussed in Sect.~\ref{s_dependence}, these are qualitatively followed by the model grid.

\subsection{Degeneracy of the solution}
\label{s_degeneracy}

\begin{figure}[tbp!]
\includegraphics[width=0.5\textwidth,angle=0]{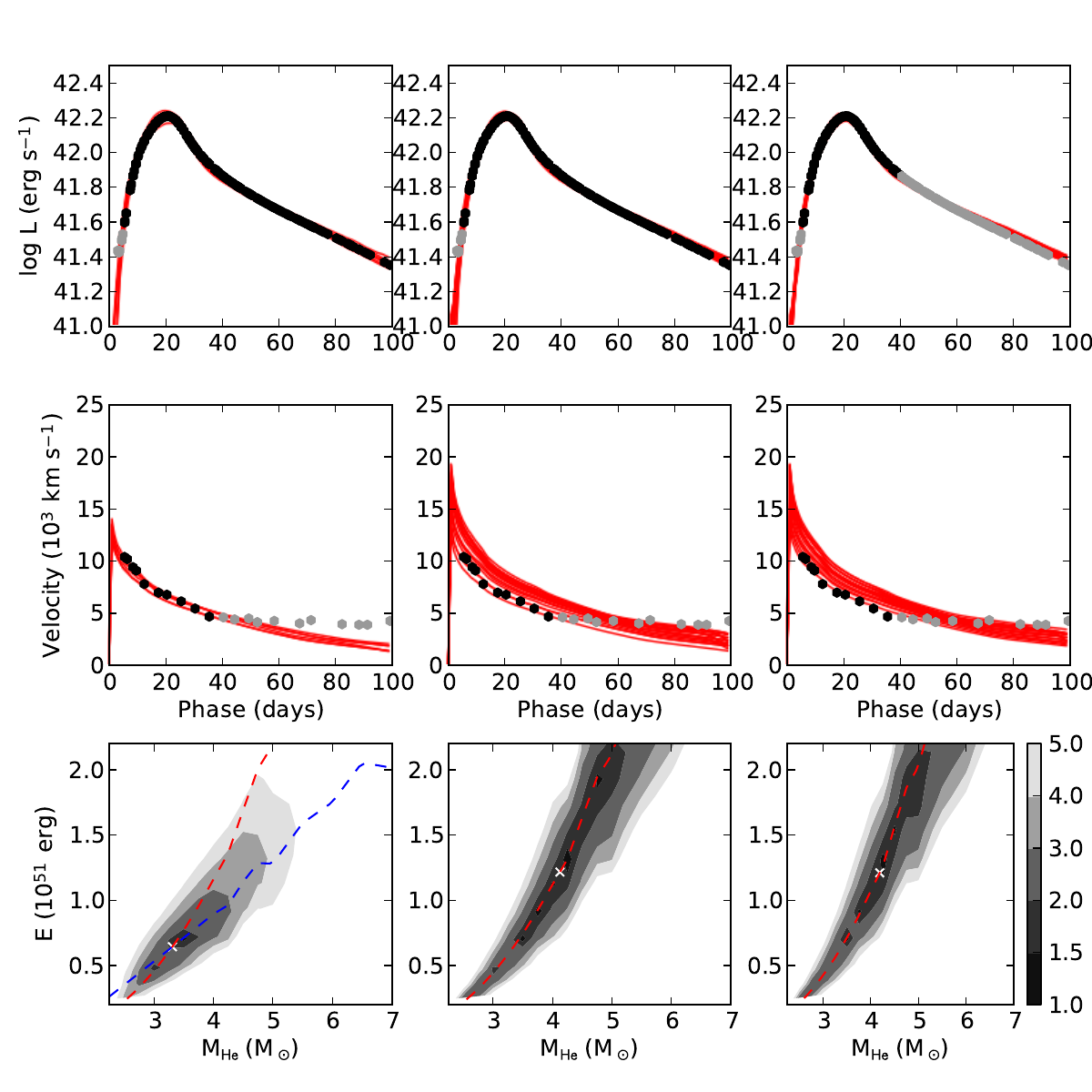}
\caption{Model bolometric lightcurve (upper panels) and photospheric velocity evolution (middle panels) as compared to the observed UV to MIR pseudo-bolometric lightcurve and estimated photospheric velocity evolution for SN 2011dh. Models with a normalized standard deviation <2 are shown in red for the cases when the lightcurve and photospheric velocity evolution (left panels), only the lightcurve (middle panels) and only the diffusion phase lightcurve (right panels) where used in the fit. The lower panels shows the corresponding contour plots displayed as in Figs.~\ref{f_mg_fit_csp_comp} and \ref{f_mg_fit_lit_comp}.}
\label{f_mg_deg}
\end{figure}

Figure~\ref{f_mg_deg} shows the bolometric lightcurve and photospheric velocity evolution for models with a normalized standard deviation <2 as compared to observations for SN 2011dh. The left, middle and right panels show the cases when lightcurve and photospheric velocity evolution, only the lightcurve and only the diffusion phase lightcurve were used in the fit, respectively. The lower panels show the corresponding contour plots and in the two latter cases, where the photospheric velocity evolution is not used in the fit, the solution is completely degenerate. This is not obvious as the diffusion phase and tail phase lightcurves might provide independent constraints arising from the diffusion time for thermal radiation and the optical depth for $\gamma$-rays, respectively. In \citetalias{Erg14a} we argued that this could be the case, given that the diffusion time and optical depth provide the constraints $\mathrm{{M_{ej}}^{3}/E_{ej}=\text{const}}$ and $\mathrm{{M_{ej}}^{2}/E_{ej}=\text{const}}$, respectively, in the \citet{Arn82} model. However, as discussed in Sect.~\ref{s_dependence}, this model assumes a constant opacity and the constraint provided by diffusion phase lightcurve is rather $\mathrm{{M_{ej}}^{2}/E_{ej}=\text{const}}$ for the hydrodynamical models. Therefore the diffusion phase and tail phase lightcurves appear to provide similar constraints and, as seen in the middle and right panels of Fig.~\ref{f_mg_deg}, the degeneracy regions well follow the $\mathrm{{M_{ej}}^{2}/E_{ej}=\text{const}}$ relation. The photospheric velocity evolution is expected to provide a constraint similar to $\mathrm{{M_{ej}}/E_{ej}=\text{const}}$, which would break the degeneracy and, as seen in the left panels of Fig.~\ref{f_mg_deg}, this is also the case. We have exemplified with SN 2011dh, but the conclusion is the same for the other SNe in the CSP and literature samples.

\subsection{Sample statistics}
\label{s_sample_statistics}

\begin{figure*}[tb!]
\includegraphics[width=1.0\textwidth,angle=0]{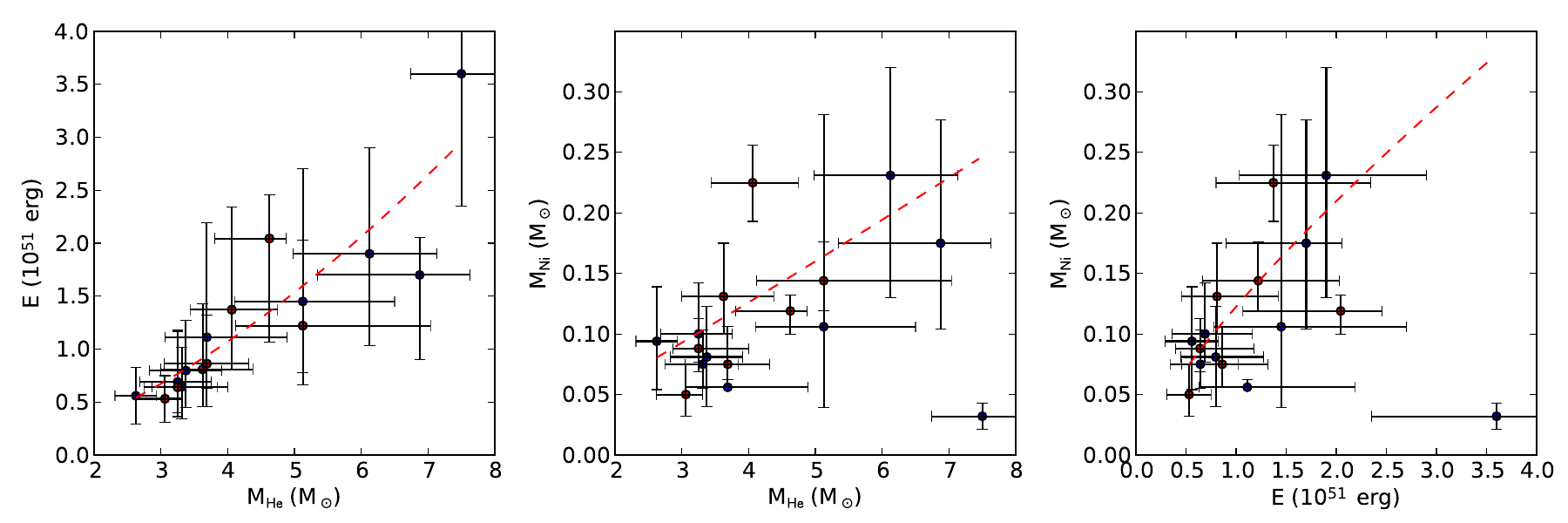}
\caption{E versus M$_{\mathrm{He}}$ (left panel), M$_{\mathrm{Ni}}$ versus M$_{\mathrm{He}}$ (middle panel) and M$_{\mathrm{Ni}}$ versus E (right panel) for our sample of Type IIb SNe (black circles), where we also show power-law fits as red dashed lines.}
\label{f_sample_stat_1}
\end{figure*}

\begin{figure}[tb!]
\includegraphics[width=0.5\textwidth,angle=0]{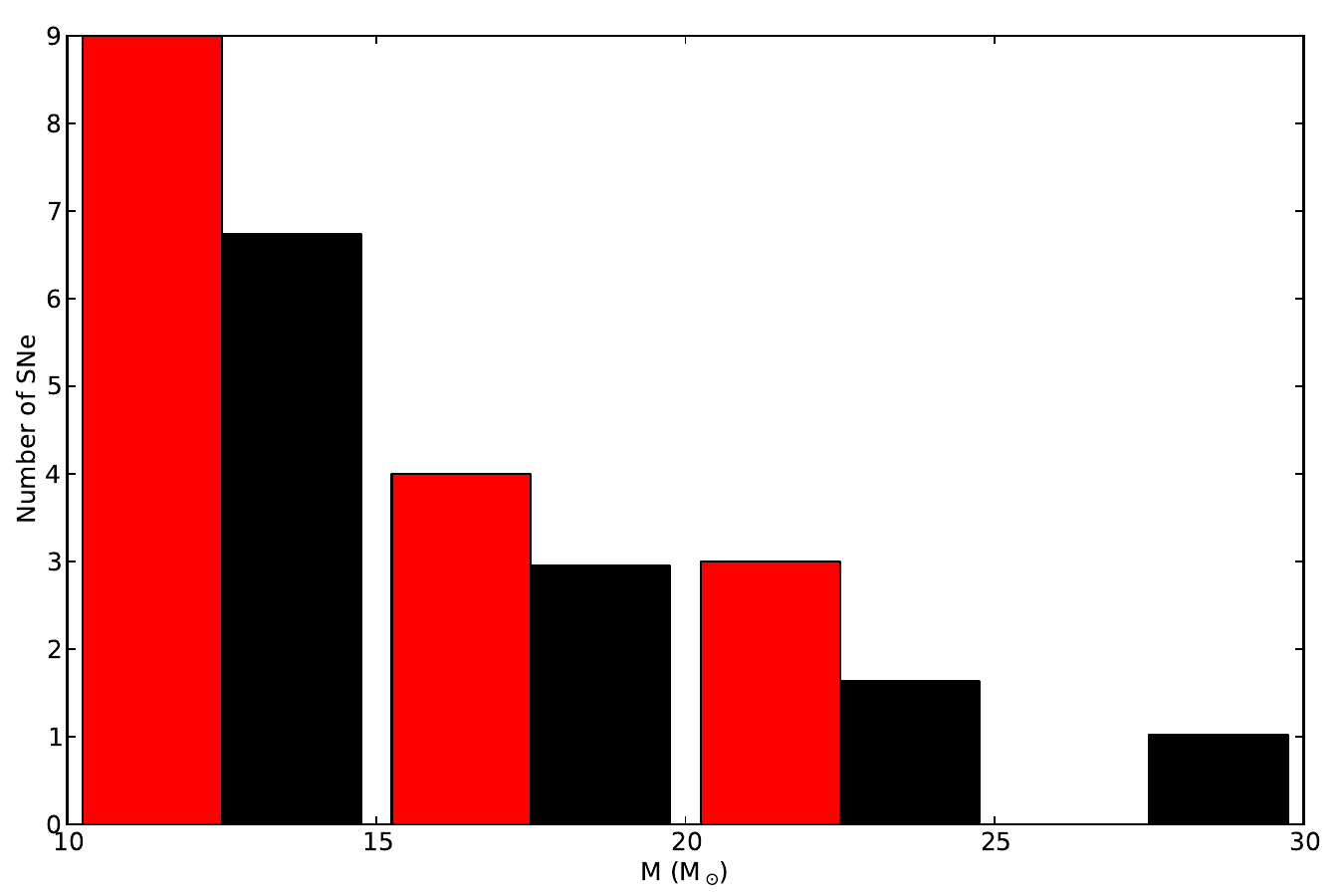}
\caption{Number of SNe with initial mass in the 8-15, 15-20, 20-25 and 25-30 M$_\odot$ bins for our sample of Type IIb SNe (red) as compared to a standard Salpeter IMF (black).}
\label{f_sample_stat_2}
\end{figure}

Our sample of Type IIb SNe, in particular those that have been individually studied in the literature, is likely biased towards odd objects, and is neither time, nor volume-limited. On the other hand, the sample is the relatively large, and the modelling used relatively advanced. No parameter studies specifically aimed at Type IIb SNe have previously been published, but the stripped-envelope SN study by \citet{Lym14} includes 8 of the Type IIb SNe in our sample, as well as SN 2004el missing from our sample. Their results are based on the approximate \citep{Arn82} model, but are nevertheless interesting to compare with. 

\subsubsection{Parameter correlations}

Figure~\ref{f_sample_stat_1} shows E versus M$_{\mathrm{He}}$ (left panel), M$_{\mathrm{Ni}}$ versus M$_{\mathrm{He}}$ (middle panel) and M$_{\mathrm{Ni}}$ versus E (right panel) for our sample of Type IIb SNe. The caveat that the explosion energy may be underestimated for SNe with relatively massive hydrogen envelopes (Sect.~\ref{s_hydrogen_effects}), and is also uncertain due to our simplified treatment of the opacity (Sect.~\ref{s_model_limitations}), should be kept in mind.

We find a correlation between E and M$_{\mathrm{He}}$, best fitted with a power law with index 1.8. Some caution is advised as this is similar to the $\mathrm{{M_{ej}}^{2}/E_{ej}=\text{const}}$ degeneracy curve (Sect.~\ref{s_degeneracy}). However, the SNe that obviously belongs to the low and high mass ends, like SNe 2011hs and 2003bg, well follows the relation. \citet{Lym14} also find a correlation, and assuming a mass for the compact remnant of 1.5 M$_\odot$ and ignoring the difference between the explosion and ejecta energy, their results for the Type IIb SNe are best fitted with a power law with index of 1.8, in agreement with our results. The correlation found is also in qualitative agreement with the relation between progenitor mass and explosion energy for Type IIP SNe suggested by \citet{Poz13}, which has a power law index of 3. 

We find correlations between M$_{\mathrm{Ni}}$ and E and between M$_{\mathrm{Ni}}$ and M$_{\mathrm{He}}$, best fitted with power laws with indices 0.7 and 1.1, respectively. \citet{Lym14} also find correlations between these quantities, and assuming a mass for the compact remnant of 1.5 M$_\odot$ and ignoring the difference between the explosion and ejecta energy, their results for the Type IIb SNe are best fitted with power laws with indices of 0.89 and 1.36, respectively, in good agreement with our results. \citet{Ham03} finds a correlation between M$_{\mathrm{Ni}}$ and E for Type IIP SNe, using relations derived from model grid fits by \citet{Lit83,Lit85}, and their results are best fitted with a power law with index 0.88, in good agreement with our results for Type IIb SNe. The correlations found are also in qualitative agreement with the relations between progenitor mass and M$_{\mathrm{Ni}}$ and expansion velocities and M$_{\mathrm{Ni}}$ for Type IIP SNe found by \citet{Fra11} and \citet{Mag12}, respectively. An increase of the mass of $^{56}$Ni with explosion energy is expected as this element is produced in the explosive nucleosynthesis, and such a trend is also predicted by modelling \citep[e.g.][]{Woo95}. We note that SN 2011ei is an extreme outlier in the M$_{\mathrm{Ni}}$ versus M$_{\mathrm{He}}$ and M$_{\mathrm{Ni}}$ versus E plots, showing the highest helium-core mass and explosion energy, but the lowest mass of $^{56}$Ni. This suggests a different nature for this SN, and a scenario where the inner parts of the ejecta fall back onto the compact remnant \citep[e.g.][]{Woo95}, strongly reducing the amount of ejected $^{56}$Ni, might be interesting to investigate.

\subsubsection{Initial mass function} 

Figure\ref{f_sample_stat_2} shows the number of SNe with initial mass in the 10-15, 15-20, 20-25 and 25-30 M$_\odot$ bins for our sample of Type IIb SNe, as well as a standard Salpeter IMF with a minimum initial mass of 10 M$_\odot$. We find no SNe instead of the expected 4.6 SNe in the >25 M$_\odot$ bins, but except for this there are no significant deviations from a standard Salpeter IMF. We find 56 percent of the SNe in the <15 M$_\odot$ bin and 81 percent of the SNe in the <20 M$_\odot$ bins. The results for the Type IIb SNe from \citet{Lym14} are similar, and assuming a mass of the compact remnant of 1.5 M$_\odot$, their results corresponds to 67 percent of the SNe in the <15 M$_\odot$ bin and 100 percent of the SNe in the <20 M$_\odot$ bins. Although single-star mass-loss rates are uncertain \citep{Smi14}, recent stellar evolutionary models \citep[e.g.][]{Eks12}, predicts the turning-point where the hydrogen envelope is lost to occur at $\sim$25 M$_\odot$ at solar metallicity for single stars, in reasonable agreement with observations of galactic RSGs \citep[e.g.][]{Lev05} and WR stars \citep[e.g.][]{Ham06}, respectively. Given this, the implication of our result is quite clear; either the binary channel is dominating the production of Type IIb SNe, or our understanding of single-star mass-loss is incomplete or even incorrect.

The idea that Type IIb SNe arise mainly from the binary channel has been proposed by several authors (references), but the size of the sample and the relatively detailed modelling used strengthen the support for this scenario considerably. Note that a similar conclusion may hold for Type Ib and Ic SNe as well, and parameter studies using approximate lightcurve modelling \citep{Can13,Tad14,Lym14} find that the ejecta masses for these SNe are typically only a few solar masses and similar to those of Type IIb SNe. As shown by \citet{Lym14} using the stellar population synthesis code BPASS \citep{Eld09}, this is considerably lower than expected for a population of single WR stars, and is also considerably lower than what would be expected for the observed population of galactic WR stars. A sample study of Type Ib and Ic SNe, using more detailed modelling, that could verify and possibly strengthen these results, would therefore be of great interest.

\section{Conclusions}
\label{s_conclusions}

We present HYDE, a new 1-D hydrodynamical code, and use it to build a grid of SN models based on bare helium-core models evolved with MESA STAR. Such a grid is well suited to model the diffusion and early tail phase of Type IIb SNe, as the progenitors of these are thought to have lost all but a tiny fraction of their hydrogen envelopes. The dependences of the observed quantities on the progenitor and SN parameters are investigated, and found to be qualitatively similar to those of the approximate \citet{Arn82} model. However, significant quantitative differences do exist, likely because of the constant opacity assumed in this model. Limitations in our method is discussed, in particular with respect to the opacity, where our simplified treatment results in a significant uncertainty in the estimated explosion energies, whereas the estimated helium-core masses are less affected. We also investigate the effects of a low-mass hydrogen envelope on the observed properties, and find these to be negligible after the luminosity minimum, expect for relatively massive hydrogen envelopes, where the photospheric velocities are decreased, likely because of deceleration of the helium core. This results in an significant uncertainty in the explosion energy for SNe with relatively massive hydrogen envelopes, and our method rather measures the explosion energy deposited in the helium core.

We use an automated fitting procedure to fit the bolometric lightcurves and photospheric velocities for our sample of Type IIb SNe to the grid of SN models. This allows us to take into account the uncertainties in distance, extinction and photospheric velocities, as well as to investigate the degeneracy of the solutions. The estimated progenitor and SN parameters for SNe 1993J, 2003bg, 2008ax, 2011dh and 2011hs are in reasonable agreement with hydrodynamical modelling in other works. However, in the case of 1993J and 2003bg, the derived explosion energies are significantly lower, likely due to the effect of their relatively massive hydrogen envelopes. We find an error in the distance and extinction to propagate mainly to derived the mass of \element[ ][56]{Ni}, and a systematic error in the photospheric velocity to propagate mainly to the derived helium-core mass and explosion energy. If the photospheric velocities are not used in the fit, there is an almost complete degeneracy along the M$_{\mathrm{ej}}^{2}$/E=const curve, but when these are included the degeneracy is broken and the fit becomes quite robust.

We find correlations between the SN and progenitor parameters, the explosion energy increasing with helium-core mass, and the mass of $^{56}$Ni increasing with the explosion energy. These correlations are best fitted with power-laws with indices of 1.8 and 1.1, respectively, in good agreement with the results in \citet{Lym14}, obtained for a smaller sample using the approximate \citet{Arn82} model. The initial masses of our sample of Type IIb SNe follows a standard Salpeter IMF reasonably well, although there is an under-population in the >25 M$_\odot$ range. The fractions of SNe with initial masses <15 M$_\odot$ and <20 M$_\odot$ are 56 and 81 percent, respectively, in good agreement with the results in \citet{Lym14}. Although single-star mass-loss rates are uncertain, single stars with initial masses much below $\sim$25 M$_\odot$ are not expected to lose their hydrogen envelopes before core-collapse, and the implication of our result is clear; either the binary channel is dominating the production of Type IIb SNe, or our understanding of single-star mass-loss needs to be revised. This conclusion is not new, and the evidence have been growing since the discovery of SN 1993J, but the size of the sample and the relatively detailed modelling used strengthen it considerably.

\section{Acknowledgements}

We thank Anders Jerkstrand for fruitful discussions on the hydrodynamical modelling, for providing the atomic data and for help with the tests of the radioactive deposition routine. We thank Melina Bersten for inspiration, for providing a tabulated version of the 4 M$_\odot$ helium core model from \citet{Nom88}, and for help with the testing of the early version of the code.

\appendix

\section{The CSP sample of Type IIb SNe}
\label{a_csp_sample}

The Type IIb SNe in the CSP sample consist of SNe 2004ex, 2004ff, 2005Q, 2006T, 2006ba, 2006bf, 2008aq, 2009K, 2009Z and 2009dq. The observational data for these SNe are described in Stritzinger~et~al.~(2015) 
and SNe 2006bf and 2009dq have been excluded due to bad sampling of the lightcurves. The photospheric velocities were estimated from the absorption minimum of the \ion{Fe}{ii} 5169 \AA~line, in turn measured from the observed spectra as described in Stritzinger~et~al.~(2015). 
Wherever used, the line-of-site extinction within the Milky way has been adopted from the \citet{Sch98} extinction maps, recalibrated by \citet{Sch11}.

\paragraph{SN 2004ex}

Discovered 2004 Oct 11.34 UT \citep{Jac04} in NGC 182 at an apparent magnitude of 17.7, and the explosion epoch is constrained by non-detections from Oct 6.35 UT (<20.0 mag) and 10.33 (<19.0 mag) \citep{Shi04}. The photometric data covers the $u$ to $H$ bands and the 5-85 days period after discovery. Both the diffusion peak and early tail are well covered. The distance modulus for NGC 182 was adopted as the median and standard deviation of the literature values given by the NASA/IPAC Extragalactic Database (NED), being 34.83$\pm{0.27}$ mag. The line-of-sight extinction within NGC 182 was estimated to $E$($B$-$V$)$_{\mathrm{H}}$=0.078 mag by comparison of the $V$-$i$ colour at $r$-band maximum with SN 2011dh. Adding the line-of-sight-extinction within the Milky Way ($E$($B$-$V$)$_\mathrm{MW}$=0.021 mag) gives $E$($B$-$V$)=0.099 mag. The pseudo-bolometric lightcurve was calculated using the $uBVriJH$ bands.

\paragraph{SN 2004ff}

Discovered 2004 Oct 30.40 UT \citep{Pug04} in ESO 552-G40 at an apparent magnitude of 18.0, and the explosion epoch is constrained by non-detections from 2004 Oct 13.41 (<19.0 mag) and 21.40 UT (<18.0 mag) \citep{Pug04}. The photometric data covers the $u$ to $H$ band and the 5-80 days period after discovery. The rise to peak is not well covered and we have included observations from \citet{Pug04} to extend this coverage. In the absence of literature measurements of the distance to ESO 552-G40 we adopt the Virgo, Great Attractor and Shapley corrected kinematic distance modulus given by NED, being 34.82$\pm{0.15}$ mag. The line-of-sight extinction within ESO 552-G40 was estimated to $E$($B$-$V$)$_{\mathrm{H}}$=0.138 mag by comparison of the $V$-$i$ colour at $r$-band maximum with SN 2011dh. Adding the line-of-sight-extinction within the Milky Way ($E$($B$-$V$)$_\mathrm{MW}$=0.029 mag) gives $E$($B$-$V$)=0.167 mag. The pseudo-bolometric lightcurve was calculated using the $uBVriJH$ bands.

\paragraph{SN 2005Q}

Discovered 2005 Jan 28.80 UT \citep{Mon05} in ESO 244-G31 at an apparent magnitude of 17.2, and the explosion epoch is constrained by a non-detection from 2004 Dec 30.81 UT (<18.7 mag). The photometric data covers the $u$ to $i$ bands and the 0-50 days period after discovery. The rise to peak and the early tail is not well covered. The distance modulus for ESO 244-G31 was adopted as the median and standard deviation of the literature values given by NED, being 34.83$\pm{0.18}$ mag. The $V$-$i$ colour at $r$-band maximum was bluer than for SN 2011dh, so the line-of-sight extinction within ESO 244-G31 was assumed to be negligible. The line-of-sight-extinction within the Milky Way is $E$($B$-$V$)$_\mathrm{MW}$=0.023 mag. The pseudo-bolometric lightcurve was calculated using the $uBVri$ bands.

\paragraph{SN 2006T}

Discovered 2006 Jan 30.99 UT \citep{Mon06a} in NGC 3054 at an apparent magnitude of 17.2, and the explosion epoch is constrained by a non-detection from 2006 Jan 16.96 UT (<18.0 mag) \citep{Mon06a}. The photometric data covers the $u$ to $H$ bands and the 0-125 days period after discovery. Both the diffusion peak and the early tail are well covered. The distance modulus for NGC 3054 was adopted as the median and standard deviation of the literature values given by NED, being 32.58$\pm{0.35}$ mag. The $V$-$i$ colour at $r$-band maximum was bluer than for SN 2011dh, so the line-of-sight extinction within NGC 3054 was assumed to be negligible. The line-of-sight-extinction within the Milky Way is $E$($B$-$V$)$_\mathrm{MW}$=0.181 mag. The pseudo-bolometric lightcurve was calculated using the $uBVriJH$ bands.

\paragraph{SN 2006ba}

Discovered 2006 Mar 19.81 UT \citep{Mon06b} in NGC 2980 at an apparent magnitude of 17.7, and the explosion epoch is constrained by a non-detection from 2006 Feb 5.04 UT (<18.8) \citep{Mon06b}. The photometric data covers the $u$ to $H$ bands and the 5-80 days period after discovery. The rise to peak is not well covered. The distance modulus for NGC 2980 was adopted as the median and standard deviation of the literature values given by NED, being 34.48$\pm{0.26}$ mag. The line-of-sight extinction within NGC 2980 was estimated to $E$($B$-$V$)$_{\mathrm{H}}$=0.212 by comparison of the $V$-$i$ colour at $r$-band maximum with SN 2011dh. Adding the line-of-sight-extinction within the Milky Way ($E$($B$-$V$)$_\mathrm{MW}$=0.046) gives $E$($B$-$V$)=0.258. The pseudo-bolometric lightcurve was calculated using the $BVriJ$ bands.

\paragraph{SN 2008aq}

Discovered 2008 Feb 27.44 UT \citep{Chu08} in MCG -02-33-20 at an apparent magnitude of 16.3, and the explosion epoch is constrained by a non-detection from 2008 Feb 10.47 UT (<19.1 mag) \citep{Chu08}. The photometric data covers the $u$ to $H$ bands and the 5-120 days period after discovery. The rise to peak is not well covered but we have included observations from \citet{Chu08} and \citet{Bro08} to extend this coverage. The distance modulus for MCG -02-33-20 was adopted as the median and standard deviation of the literature values given by NED, being 32.45$\pm{0.43,0.43}$ mag. The $V$-$i$ colour at $r$-band maximum was bluer than for SN 2011dh so the line-of-sight extinction within MCG -02-33-20 was assumed to be negligible. The line-of-sight-extinction within the Milky Way is $E$($B$-$V$)$_\mathrm{MW}$=0.040 mag. The pseudo-bolometric lightcurve was calculated using the $uBVriJH$ bands.

\paragraph{SN 2009K}

Discovered 2009 Jan 14.07 UT \citep{Pig09} in NGC 1620 at an apparent magnitude of 14.9, and the explosion epoch is constrained by a non-detection from 2009 Jan 11.08 UT (<18.0 mag) \citep{Pig09}. The photometric data covers the $B$ to $H$ bands and the 0-50 days period after discovery. The early tail is not covered. The distance modulus for NGC 1620 was adopted as the median and standard deviation of the literature values given by NED, being 33.15$\pm{0.22}$ mag. The line-of-sight extinction within NGC 1620 was estimated to $E$($B$-$V$)$_{\mathrm{H}}$=0.057 mag by comparison of the $V$-$i$ colour at $r$-band maximum with SN 2011dh. Adding the line-of-sight-extinction within the Milky Way ($E$($B$-$V$)$_\mathrm{MW}$=0.051 mag) gives $E$($B$-$V$)=0.108 mag. The pseudo-bolometric lightcurve was calculated using the $uBVri$ bands.

\paragraph{SN 2009Z}

Discovered 2009 Feb 2.53 UT \citep{Gri09} in SDSS J140153.80-012035.5 at an apparent magnitude of 18.1, and the explosion epoch is constrained by a non-detection from 2008 Jun-Jul (<19.4 mag) \citep{Gri09}. The photometric data covers $u$ to $i$ bands and the 5-85 days period after discovery, although additional NIR photometry was obtained at $\sim$400 days. The rise to peak is not well covered but we have included observations from \citet{Gri09} to extend this coverage. In the absence of literature measurements of the distance to SDSS J140153.80-012035.5 we adopt the the Virgo, Great Attractor and Shapley corrected kinematic distance modulus given by NED, being 35.26$\pm{0.15}$ mag. The $V$-$i$ colour at $r$-band maximum was bluer than for SN 2011dh so the line-of-sight extinction within SDSS J140153.80-012035.5 was assumed to be negligible. The line-of-sight-extinction within the Milky Way is $E$($B$-$V$)$_\mathrm{MW}$=0.042 mag. The pseudo-bolometric lightcurve was calculated using the $uBVri$ bands.

\section{Individually studied Type IIb SNe}
\label{a_literature_sample}

The sample of Type IIb SNe that have been individually studied in the literature consist of SNe 1993J \citep[e.g.][]{Ric94,Ric96}, 1996cb \citep{Qiu99}, 2003bg \citep{Hamu09}, 2008ax \citep[e.g.][]{Tau11}, 2011dh \citepalias[e.g.][]{Erg14a}, 2011fu \citep{Kum13}, 2011hs \citep{Buf14}, 2011ei \citep{Mil13} and 2013df \citep{Dyk14}, omitting here 2010as \citep{Fol14b}, for which the data was published after our work began. Wherever used, the line-of-site extinction within the Milky way has been adopted from the \citet{Sch98} extinction maps, recalibrated by \citet{Sch11}.

\paragraph{SN 1996cb}

Discovered 1996 Dec 15.71 UT \citep{Nak96} in NGC 3510 at an apparent magnitude of 16.5, and the explosion epoch is constrained by a non-detection from 1996 Nov 29 UT (<19.0 mag) \citep{Qia96}. The photometric data was taken from \citet{Qiu99} and covers the $B$ to $R$ bands and the 5-160 days period after discovery. To extend the rise to peak coverage we also included observations from \citep{Nak96} and \citep{Qia96}. The distance modulus for NGC 3510 was adopted as the mean and standard deviation of the literature values given by NED being 30.57$\pm{1.03}$. The total line-of-sight extinction was taken as the mean of the upper limit determined by \citet{Qiu99} from comparison to SN 1993J ($E$($B$-$V$)=0.12) and the line-of-sight extinction within the Milky Way ($E$($B$-$V$)=0.12), giving $E$($B$-$V$)=0.073$\pm{0.047}$. The pseudo-bolometric lightcurve was calculated using the $BVR$ bands. Estimates of the photospheric velocities using SYNOW were adopted from \citet{Den01}. \citet{Tak12} showed that this method gives results similar to those obtained from the absorption minimum of \ion{Fe}{ii} 5169 \AA~line for a sample of Type IIP SNe.

\paragraph{SN 2003bg}

Discovered 2003 Feb 25.70 UT \citep{Woo03} in MCG -05-10-15 at an apparent magnitude of 15.0, and the explosion epoch is constrained by a non-detection from 2002 Nov 7.0 UT (<18.0 mag) \citep{Woo03}. The photometric data was taken from \citet{Hamu09} and covers the $B$ to $K$ bands and the 5-325 days period after discovery. The distance modulus for the host galaxy MCG-05-10-015 was adopted from \citet{Kel00}, and the total line-of-sight extinction taken as the mean of the line-of-sight extinction within the Milky Way ($E$($B$-$V$)=0.02) and an assumed upper limit of 0.1 mag additional extinction, giving $E$($B$-$V$)=0.070$\pm{0.05}$. The pseudo-bolometric lightcurve was calculated using the $BVRIJHK$ bands. Estimates of the photospheric velocities using the Monte Carlo (MC) radiative transfer code by \citet{Maz93,Luc99,Maz00} was adopted from \citet{Maz09}.

\paragraph{SN 2011ei}

Discovered 2011 Jul 25.43 UT \citep{Mar11b} in NGC 6925 at an apparent magnitude of 18.0, and the explosion epoch is constrained by a non-detection from 2011 Jun 23.58 UT (<19.1 mag) \citep{Mar11b}. The photometric data was taken from \citet{Mil13} and covers $U$ to $I$ bands and the 0-50 days period. The distance modulus for the host galaxy NGC 6925 was adopted as the mean and standard deviation of the literature values given by NED, being 32.42$\pm{0.27}$. The total line-of-sight extinction was taken as the mean of the upper limit determined by \citet{Mil13} ($E$($B$-$V$)=0.232), who used the equivalent width of the interstellar \ion{Na}{i} D interstellar absorption lines and the relation between this and $E$($B$-$V$) by \citet{Tur03}, and the line-of-sight-extinction within the Milky Way ($E$($B$-$V$)=0.052), giving $E$($B$-$V$)=0.142$\pm{0.09}$. The pseudo-bolometric lightcurve was calculated using the $UBVRI$ bands. Estimates of the photospheric velocities using SYNOW was taken from \citet{Mil13}.

\paragraph{SN 2011fu}

Discovered 2011 Sep 21.04 UT \citep{Cia11} in UGC 1626 at an apparent magnitude of 15.8, and the explosion epoch is constrained by a non-detection from 2011 Aug 10 UT (<18.8 mag) \citep{Cia11}. The photometric data covers the $U$ to $I$ bands and the 10-175 days period and was taken from \citet{Kum13}. In the absence of literature measurements of the distance to the host galaxy UGC 1626 we adopt the Virgo, Great Attractor and Shapley corrected kinematic distance modulus given by NED, being 34.36, and assume an error in this estimate of 50 percent. The total line-of-sight extinction was adopted from \citet{Kum13}, who used the equivalent width of the interstellar \ion{Na}{i} D interstellar absorption lines and the relation between this and $E$($B$-$V$) by \citet{Mun97} to estimate $E$($B$-$V$)=0.22$\pm{0.11}$. The pseudo-bolometric lightcurve was calculated using the $UBVRI$ bands. Measurements of the absorption minimum of the \ion{Fe}{ii} 5169 \AA~line was taken from \citet{Kum13}. 

\paragraph{SN 2011hs}

Discovered 2011 Nov 12.48 UT \citep{Dre11} in IC 5267 at an apparent magnitude of 15.5, and the explosion epoch is constrained by a non-detection from 2011 Oct 9.574 UT (<18.7 mag) \citep{Dre11}. The photometric data covers the $U$ to $K$ bands and 0-120 days period and was taken from \citet{Buf14}. The distance modulus for the host galaxy IC 5267 was adopted as the mean and standard deviation of the literature values given by NED, being 32.18$\pm{0.41}$. The total line-of-sight extinction was estimated to $E$($B$-$V$)$_{\mathrm{H}}$=0.400 by comparison of the $V$-$i$ colour at $r$-band maximum with SN 2011dh.
Spectroscopic data was taken from \citet{Buf14}, and the photospheric velocities estimated from the absorption minimum of the \ion{Fe}{ii} 5169 \AA~line, in turn measured with a simple automated algorithm described in \citetalias{Erg14a}.

\paragraph{SN 2013df}

Discovered 2013 Jun 7.87 UT \citep{Cia13} in NGC 4414.
The photometric data covers the $B$ to $H$ bands and the 5-65 days period and was taken from \citet{Dyk14}. The distance modulus for the host galaxy NGC 4414 was taken as the Cepheid based measurement by \citet{Fre01}, being 31.10$\pm{0.05}$. The total line-of-sight extinction was adopted from \citet{Dyk14}, which used the equivalent width of the interstellar \ion{Na}{i} D interstellar absorption lines and the relation between this and $E$($B$-$V$) by \citet{Poz12} to estimate $E$($B$-$V$)=0.097$\pm{0.016}$. The pseudo-bolometric lightcurve was calculated using the $BVRIzJH$ bands. The spectroscopic data was taken from \citet{Dyk14} and the photospheric velocities estimated from the absorption minimum of the \ion{Fe}{ii} 5169 \AA~line in turn measured with a simple automated algorithm described in \citetalias{Erg14a}.

\bibliographystyle{aa}
\bibliography{hydro-IIb}

\label{lastpage}

\end{document}

%% file: mg-mesa-model-table.tex
\begin{tabular}{lllll}
\hline\hline \\ [-1.5ex]
M$_{\mathrm{He}}$ & M$_{\mathrm{C/O}}$ & M$_{\mathrm{Fe}}$ & R & E$_\star$ \\ [0.5ex]
(M$_\odot$) & (M$_\odot$) & (M$_\odot$) & (R$_\odot$) & (10$^{51}$ erg) \\
\hline \\ [-1.5ex]
4.00 & 2.04 & 1.32 & 4.81 & -0.19\\
4.25 & 2.21 & 1.46 & 4.21 & -0.19\\
4.50 & 2.39 & 1.57 & 3.51 & -0.30\\
4.75 & 2.58 & 1.61 & 4.22 & -0.37\\
5.00 & 2.75 & 1.56 & 4.19 & -0.30\\
5.50 & 3.09 & 1.66 & 3.75 & -0.47\\
6.00 & 3.54 & 1.69 & 3.97 & -0.49\\
6.50 & 3.92 & 1.77 & 4.13 & -0.65\\
7.00 & 4.32 & 1.51 & 3.96 & -0.47\\ [0.5ex]
\hline
\end{tabular}

%% file: mg-fit-csp-comp-table.tex
\begin{tabular}{llllll}
\hline\hline \\ [-1.5ex]
SN & E & M$_{\mathrm{He}}$ & M$_{\mathrm{Ni}}$ & Mix$_{\mathrm{Ni}}$ & JD$_{\mathrm{exp}}$ (+2400000)\\ [0.5ex]
& (10$^{51}$ erg) & (M$_\odot$) & (M$_\odot$) & & (days) \\
\hline \\ [-1.5ex]
2004ex & 0.81 (+0.62,-0.36) & 3.62 (+0.75,-0.62) & 0.131 (+0.044,-0.031) & 1.00 (+0.21,-0.10) & 53285.84 (+0.50,-0.56)\\
2004ff & 2.04 (+0.41,-0.98) & 4.62 (+0.25,-0.82) & 0.119 (+0.013,-0.019) & 1.55 (+0.05,-0.76) & 53294.89 (+1.80,-0.50)\\
2005Q & 1.03 (+0.92,-0.38) & 4.00 (+1.02,-0.62) & 0.181 (+0.038,-0.026) & 1.40 (+0.00,-0.49) & 53385.55 (+2.92,-1.00)\\
2006T & 0.86 (+0.46,-0.41) & 3.69 (+0.63,-0.64) & 0.075 (+0.031,-0.019) & 1.20 (+0.40,-0.10) & 53762.96 (+0.50,-1.03)\\
2006ba & 0.64 (+0.54,-0.24) & 3.25 (+0.75,-0.38) & 0.088 (+0.025,-0.019) & 1.60 (+0.00,-0.21) & 53805.31 (+1.00,-0.50)\\
2008aq & 0.53 (+0.22,-0.22) & 3.06 (+0.25,-0.44) & 0.050 (+0.025,-0.018) & 1.20 (+0.00,-0.55) & 54514.94 (+2.24,-0.00)\\
2009K & 1.22 (+0.81,-0.56) & 5.12 (+1.91,-1.01) & 0.144 (+0.032,-0.025) & 1.12 (+0.18,-0.08) & 54844.08 (+0.00,-0.00)\\
2009Z & 1.37 (+0.97,-0.57) & 4.06 (+0.69,-0.62) & 0.225 (+0.031,-0.032) & 1.60 (+0.00,-0.05) & 54859.53 (+0.50,-0.00)\\
\hline
\end{tabular}

%% file: mg-fit-lit-comp-table.tex
\begin{tabular}{llllll}
\hline\hline \\ [-1.5ex]
SN & E & M$_{\mathrm{He}}$ & M$_{\mathrm{Ni}}$ & Mix$_{\mathrm{Ni}}$ & JD$_{\mathrm{exp}}$ (+2400000)\\ [0.5ex]
& (10$^{51}$ erg) & (M$_\odot$) & (M$_\odot$) & & (days) \\
\hline \\ [-1.5ex]
1993J & 0.69 (+0.47,-0.33) & 3.25 (+0.50,-0.57) & 0.100 (+0.042,-0.023) & 0.85 (+0.14,-0.07) & ...\\
1996cb & 1.45 (+1.25,-0.67) & 5.12 (+1.38,-1.02) & 0.106 (+0.175,-0.067) & ... & 50430.50 (+0.71,-1.12)\\
2003bg & 1.70 (+0.36,-0.80) & 6.88 (+0.75,-1.53) & 0.175 (+0.102,-0.071) & ... & 52689.50 (+0.00,-0.00)\\
2008ax & 0.80 (+0.48,-0.35) & 3.38 (+0.53,-0.55) & 0.081 (+0.042,-0.041) & 0.90 (+0.00,-0.11) & ...\\
2011dh & 0.64 (+0.38,-0.30) & 3.31 (+0.54,-0.57) & 0.075 (+0.028,-0.020) & 1.05 (+0.08,-0.00) & ...\\
2011ei & 3.60 (+2.44,-1.25) & 7.50 (+2.00,-0.76) & 0.032 (+0.011,-0.011) & ... & 55763.00 (+1.12,-0.00)\\
2011fu & 1.90 (+1.00,-0.87) & 6.12 (+1.01,-1.15) & 0.231 (+0.089,-0.101) & ... & 55820.50 (+1.12,-3.84)\\
2011hs & 0.56 (+0.27,-0.27) & 2.62 (+0.31,-0.31) & 0.094 (+0.045,-0.040) & 1.55 (+0.05,-0.78) & 55874.50 (+2.55,-0.00)\\
2013df & 1.11 (+1.08,-0.48) & 3.69 (+1.20,-0.63) & 0.056 (+0.006,-0.000) & 0.80 (+0.60,-0.16) & 56449.80 (+2.12,-0.00)\\
\hline
\end{tabular}